\documentclass[9pt,twocolumn,twoside]{pnas-new}

\usepackage[modulo]{lineno}

\articletype{Astronomy}

\templatetype{pnasresearcharticle} 

\begin{document}
\nolinenumbers

\title{Modified Newtonian Dynamics and galaxy clusters}

\author[a]{Francoise Combes}

\affil[a]{Observatoire de Paris, LUX (Laboratoire d’étude de l’Univers et des phénomènes eXtrêmes), Collège de France, CNRS, Paris-Sciences-Lettres University, Sorbonne University, 75014, Paris, France}

\leadauthor{Combes}

\significancestatement{One of the paramount enigmas in astrophysics is dark matter, an elusive
hypothetical substance representing 25\% of the Universe. Dark matter
remains invisible, detected only by its gravitational influence on visible
structures: from governing the formation and rotation of galaxies to
inducing strong and weak gravitational lensing in galaxy clusters.
While the standard cosmological model assumes the existence of
exotic Cold Dark Matter (CDM) particles, their ongoing
non-detection, combined with the CDM model’s challenges at galactic
Scales, has attracted interest in Milgrom’s Modified Newtonian
Dynamics (MOND). However, while MOND remarkably reproduces galaxy
rotation dynamics, it fails when applied to cluster scales. This paper
explores whether undetected ordinary matter harbored within galaxy clusters
could solve the problem.
}

\authorcontributions{F.C. performed the research and wrote the paper}
\authordeclaration{No competing interests}
\correspondingauthor{\textsuperscript{2} E-mail: francoise.combes@obspm.fr}


\begin{abstract}
First proposed by Milgrom in 1983 as an alternative to the standard Cold Dark Matter (CDM) paradigm, Modified Newtonian Dynamics (MOND) has gained renewed momentum in light of persistent non-detections of exotic dark matter particles and growing tensions facing CDM on galactic scales. While MOND perfectly reproduces galactic rotation curves, it encounters notable challenges in explaining the equilibrium of galaxy clusters. Although it reduces the overall dark matter budget in these massive systems, an unidentified mass component remains required.
This paper investigates the hypothesis that this missing mass is baryonic: composed of non-radiating ordinary matter such as stellar remnants, gas too diffuse to initiate star formation, or compact, ultra-dense clouds of cold molecular gas with low volume filling factors. Using a sample of 106 clusters mapped in X-rays, emitted by the hot intra-cluster medium (ICM) composing most of the visible mass, we derive first the total mass within the Newtonian dynamics, assuming hydrostatic equilibrium and spherical symmetry. Then, with the modified Poisson equation, we reconstruct the spatial distribution of the invisible missing matter within the MOND framework.
In the sample, the relaxed systems harboring cooling flows, known as cool-core clusters, have the highest peaks of dark missing mass. Its density profile reveals a pronounced central concentration bounded by a steep truncation immediately beyond a core radius of 20–30 kpc, confirming previous studies. MOND reduces the necessary dynamical mass by a factor of 4 to 10.  The residual missing mass is proportional to the total gas mass, being roughly twice lower.
\end{abstract}

\dates{This manuscript was compiled on \today}
\doi{\url{www.pnas.org/cgi/doi/10.1073/pnas.XXXXXXXXXX}}

\maketitle
\ifthenelse{\boolean{shortarticle}}{\ifthenelse{\boolean{singlecolumn}}{\abscontentformatted}{\abscontent}}{}

\Firstpage

The Modified Newtonian Dynamics (MOND) is an empirical model proposed by Milgrom in 1983 \cite{Milgrom1983ApJ...270..365M}
as an alternative to assuming the existence of hidden dark matter in the cosmos. The model 
suggests a modified gravity law at weak accelerations, close or below 
a$_0 = 1.2 \times 10^{-10}$m/s$^2$. The model succeeds to fit most galaxy 
rotation curves without dark matter \cite[e.g.][]{Lelli2016AJ....152..157L}. It explains naturally galaxy scaling relations, such as the baryonic Tully-Fisher or Faber-Jackson, which are only reproduced in a contrived fashion
in the standard $\Lambda$CDM model \cite{Famaey2012LRR....15...10F}.  If the MOND model solves some puzzles associated to the CDM model, such as the predicted density cusp in the center of galaxies, while a core is observed, or the numerous predicted galaxy satellites, missing in observations, it also has some problems of its own. MOND reduces the required dark matter in galaxy clusters, but is not sufficient to remove it completely: The \& White in 1988 
\cite{The1988AJ.....95.1642T} remarked that it was possible to remove it in the Coma cluster,
only when the critical acceleration a$_0$ was multiplied by 3. 
 Colliding clusters are an occasion to separate baryons from dark matter and were claimed to falsify MOND
\cite{Clowe2004ApJ...604..596C}.
However, baryons and dark matter are separated in galaxies, and the Bullet Cluster
is no more a problem than any other galaxy cluster \cite{Hernandez2026ApJ..1008L..46H}.

Extending this analysis to a sample of X-ray-emitting clusters, Sanders (1999) \cite{Sanders1999ApJ...512L..23S} concluded that the ratio of the dynamical to observable mass is only reduced by a factor two when MOND gravity is considered.

%

Several approaches have been proposed to resolve this residual mass discrepancy in galaxy clusters under MOND.

\Endparasplit
Notably, Angus and collaborators \cite{Angus2008MNRAS.387.1470A, Angus2009MNRAS.394..527A} investigated the inclusion of 11 eV sterile neutrinos as hot dark matter. Although a viable mechanism, this hybrid scheme compromises MOND’s primary appeal: dispensing with dark matter altogether. Nevertheless, it has proven valuable for probing galaxy formation and evolution within a cosmological context. Pending a fully fledged relativistic MOND theory capable of underpinning cosmological simulations, this $\nu$HDM model has served as a realistic ansatz for running structure-formation simulations \cite{Samaras2025MNRAS.541..784S}.

A plausible candidate for the missing mass in galaxy clusters is dark baryons. This hypothesis was originally advanced by Milgrom (2008) \cite{Milgrom2008NewAR..51..906M}, who noted that these dark baryons ought to be concentrated within cluster cores, effectively doubling the clusters' total baryonic mass. Because the collapsed mass fraction of the Universe, including galaxy groups and clusters, accounts for merely 20\% of the total mass budget \cite{Cautun2014MNRAS.441.2923C}, there is sufficient latitude to accommodate an increased cluster mass. Clusters harbor less than 10\% of galaxies.
The majority of galaxies belong to groups, an environment where MOND succeeds quite well. The gravity field is weak enough in galaxy groups to be all over the place 1-2 orders of magnitude below a$_0$ \cite{Milgrom2018PhRvD..98j4036M}.

Furthermore, a complete census of cosmic baryons remains elusive. Galaxies account for a mere 7\% of the total baryon budget of the Universe, while the hot intracluster gas contributes only 4\% \cite{Shull2012ApJ...759...23S}. Roughly 20\% resides in the Lyman-$\alpha$ forest within cosmic filaments, traced via its minute neutral fraction. The Warm-Hot Intergalactic Medium (WHIM) is estimated with large uncertainty to harbor an additional 30–40\%, leaving at least 30\% of the Universe's baryons unaccounted for \cite{Migkas2025A&A...698A.270M, Gatuzz2023MNRAS.521.3098G,  Bonamente2025MNRAS.542....1B}.

\section{Mass distribution required for dark baryons in clusters}


In MOND phenomenology, all matter is assumed to be baryonic. Within the deep MOND regime (DMR), where acceleration falls below the critical threshold $a_0$, the radial acceleration $a = V^2/R$ asymptotically approaches $\sqrt{a_N a_0}$, with $a_N$ denoting the Newtonian acceleration, and $V$ the rotational velocity. From this relation, the baryonic Tully-Fisher relation (M$ \propto V^4$) emerges naturally. Non-rotating systems, such as elliptical galaxies, obey an equivalent scaling law via the Faber-Jackson relation, substituting velocity dispersion $\sigma$ for rotational velocity $V$. Notably, this framework extends to galaxy clusters, as first demonstrated by Sanders (1994) \cite{Sanders1994A&A...284L..31S}. As highlighted by Famaey \& McGaugh (2012, Fig. 48) \cite{Famaey2012LRR....15...10F}, cluster data align even more tightly when assuming a baryonic mass roughly twice that of the observed visible matter; this suggests that these scaling relations may serve as a valuable tool for tracing missing baryons.

Note that the MOND model is not limited by the universal ratio between baryons and dark matter, like in the standard CDM model. In some rich clusters, the fraction of baryons tends to reach the universal value, and doubling the baryonic mass is impossible in $\Lambda$CDM cosmology.

Within galaxy clusters, the missing mass problem is localized primarily in the central core, a region that operates within the Newtonian regime, where gravitational acceleration is not small relative to the critical scale a$_0$. This points to the plausible existence of dark baryons concentrated in cluster cores, packed even more densely than traditional cold dark matter models assume. Such high central concentration, followed with a steep radial cutoff, is vital to avoid overpredicting the gravitational force in the cluster's outer regions.

Because the isothermal-sphere density profile in MOND typically decays steeply as r$^{-4}$, accommodating the observed cluster dynamics requires a two-component baryonic model \cite{Milgrom1984ApJ...287..571M, Gerbal1992A&A...262..395G,Sanders2003MNRAS.342..901S}. Farther out in the cluster periphery, local gravitational accelerations fall well below external environmental interactions, making it necessary to account for the External Field Effect (EFE). Under the influence of the EFE, the gravitational force law recovers its standard r$^{-2}$ behavior, where the visible mass of the hot intracluster medium (ICM) becomes sufficient to explain the observed kinematics. However, mass density profiles at large radii remain subject to observational uncertainty, leaving the validity of hydrostatic equilibrium open to debate.

A variety of observational methods have attempted to map this missing baryonic component. Examining X-ray observations for more than forty clusters under the assumption of hydrostatic equilibrium, Sanders (2003) \cite{Sanders2003MNRAS.342..901S} inferred an unaccounted mass averaging 1.6 times the ICM mass, though anomalies remained in the outer regions. Later, Kelleher \& Lelli (2024) \cite{Kelleher2024A&A...688A..78K} analyzed six X-COP clusters using combined X-ray, kinematics, and Sunyaev–Zel'dovich data; they likewise identified a missing mass comparable or slightly higher than the ICM mass, but successfully resolved the outer boundary dynamics by incorporating the EFE. Reinforcing this picture, gravitational lensing studies across sixteen and twenty clusters yielded consistent results, demonstrating a densely packed core distribution characterized by a sharp truncation near $\sim$ 400 kpc \cite{Tian2020ApJ...896...70T, Famaey2025PhRvD.111l3042F}. Consequently, while the total hidden mass remains comparable to that of the hot gas, the extreme central compaction means the surface density within the core could exceed that of the hot gas by a factor 10.

\begin{table}[t!] 
\centering
\caption{Previous MOND studies of galaxy clusters} \label{Tab:previous}
\begin{tabular}{lrrrr}
Authors& Clusters & z     & Mass         & CC  \\
       & Number   &median & E14 M$_\odot$& number    \\
\midrule
Sanders (2003)$^1$           &42 & 0.055 & 7.80 & 24 \\
Tian et al. (2020)$^2$       &20 & 0.375 & 7.72 & 10 \\
Kelleher \& Lelli (2024)$^3$ & 5 & 0.070 & 7.80 &  2 \\
Famaey et al (2025)$^2$      &16 & 0.392 & 6.55 &  7 \\
\bottomrule
\end{tabular}

\addtabletext{Clusters from $^1$ HIFLUGCS $^2$ CLASH 
(Cluster Lensing And Supernova survey with Hubble) $^3$ X-COP (Cluster Outskirt Project)}
\end{table}

\section{New computation with more than 100 nearby clusters}
 
Prior studies have investigated galaxy-cluster dynamics within the MOND framework to constrain the distribution of missing mass. These studies examined clusters across varying redshifts, both local and distant, utilizing diverse total mass estimators, including galaxy kinematics, gravitational lensing, and the hydrostatic equilibrium of the hot ICM gas mapped via X-ray and Sunyaev-Zel’dovich observations. Sample sizes and median parameters from these investigations are summarized in Table \ref{Tab:previous}. Building on this foundation, our objective is to enhance statistical robustness by analyzing the extended HIFLUGCS sample of 106 clusters \cite{Chen2007A&A...466..805C}, enabling a clear distinction between Cool-Core (CC) and Non-Cool-Core (Non-CC) populations. These two regimes likely represent distinct evolutionary phases: while cluster mergers intermittently disrupt dynamical equilibrium, intervening quiescent periods allow the gas to cool, forming "dark" baryons that remain dispersed throughout the multiphase ICM.

\subsection{The cluster sample and their dynamical mass}
We use the extended HIFLUGCS (the Highest X-ray FLUx Galaxy Cluster Sample)  containing  106 clusters and groups of galaxies, selected only from X-rays. They belong to the nearby Universe, their median redshift being 0.048 \cite{Reiprich2002ApJ...567..716R, Chen2007A&A...466..805C}. Most clusters are quasi isothermal and characterized by their temperature $T_h$, with a median value of 4.5 keV. In less than half the sample, the temperature slightly decreases in the center, revealing a cooling core (CC). However, the hot gas involved in this component has a tiny contribution to the mass. The gravitational potential of the cluster and its total mass are reflected in the $T_h$ component.
The hot gas of the ICM is detected up to R$_{500}$, the radius at which the mean density of the cluster is 500 times that of the critical density universe. 
In CC clusters, it is possible to compute the cooling time-scale, smaller than a Hubble time,
and the expected cooling flow rate $\dot{M}$ \cite{Chen2007A&A...466..805C}. The median value 60
M$_\odot$ yr$^{-1}$ separates the weak-CC from the CC clusters, cf. Table \ref{Tab:CC}.

The ICM density is represented by the widely used $\beta$-model, with a density
\begin{equation}\label{eqn:betamodel}
\rho =\rho_0 (1 + (r/r_c)^2)^{-3\beta/2}
\end{equation}
 with $r_c$ the core radius. The latter is the turn over radius, from an almost flat density towards a steep power-law in the outer parts. Note that the value $\beta$ =1 delimits the slopes for which the mass converges to a finite value at large radii. The median value for the sample is $\beta$ = 0.63.
The core radius of the CC clusters is about tens of kpc (around 30~kpc), while for non-CC clusters, it is of the order of hundreds of kpc, around 300~kpc \cite{Chen2007A&A...466..805C}. This strong difference might be the result of evolution and relaxation, and will teach us about the distribution of dark mass.
Assuming spherical symmetry and the gas in hydrostatic equilibrium, the gravitational acceleration is $g = 1/\rho$ $dP/dr$, with the pressure $P = \rho  c_s^2$. In Newtonian gravity, the total mass of the cluster can be calculated for this $\beta $ model as
\begin{equation}\label{eqn:Mdyn}
M(<r)=\frac{3\beta c_s^2 r}{G}\frac{(r/r_c)^2}{1+(r/r_c)^2} ,
\end{equation}
where $G$ is the gravitational constant,
and $c_s$ the velocity of sound in the hot medium verifies
\begin{equation}\label{eqn:sound-speed}
c_s^2 = \frac{kT_h}{\mu m_p}
\end{equation}
where $k$ is the Boltzmann constant, $\mu$ the molecular weight ($\mu$=0.61), and $m_p$ the proton mass.

 Two gas components have been fitted for 49 clusters by \cite{Chen2007A&A...466..805C}, from their X-ray emission where the surface density is proportional to the square of the density. The two components are modeled as in Eq. \ref{eqn:betamodel}, with $\rho_{01}$ and $\rho_{02}$  their central density, $r_{c1}$ and  $r_{c2}$ their core radii, and $\beta_1$,  $\beta_2$ their exponents.
The total central density, $\rho_0$, can then be derived from
\begin{equation}
\rho(r)^2=\rho_{01}^2 \alpha_1^{-3\beta_1}+\rho_{02}^2 \alpha_2^{-3\beta_2}
\end{equation}
with the help of the following notation:
\begin{equation}
\alpha_1=1+(r/r_{c1})^2 ,  \alpha_2=1+(r/r_{c2})^2 ,
\end{equation}
It is possible to compute the dynamical mass as in Eq. \ref{eqn:Mdyn} by
\begin{equation}\label{eqn:Mdyn2beta}
M(<r)= c_s^2 \frac{3r^3}{G}\frac{\rho_{01}^2 \beta_1\alpha_1^{-3\beta_1-1}/r_{c1}^2
     +\rho_{02}^2 \beta_2\alpha_2^{-3\beta_2-1}/r_{c2}^2}
     {\rho_{01}^2\alpha_1^{-3\beta_1}+\rho_{02}^2\alpha_2^{-3\beta_2}} 
\end{equation}

When variations of the gas temperature $T_{\rm gas}(r)$ are taken into account:
\begin{equation}
M(<r) = - \frac{kT_{\rm gas}(r) \, r}{\mu m_p G} 
 \left( \frac{\partial \log T_{\rm gas}}{\partial \log r} +
   \frac{\partial \log n_{\rm gas}}{\partial \log r} \right).
\label{eq:Tvary}
\end{equation}
  We have taken into account the positive temperature gradient in the core of the CC clusters. This was modelled by
 a linear increase with radius, with the central gas temperature 20\% lower than the global T$_h$ of the cluster.
 It is not physical to take into account lower temperatures, since then no hydrostatic equilibrium is possible. 

\subsection{Computation of the missing mass}

\subsubsection{The mass required by MOND in clusters}
\begin{figure}
\centering
\includegraphics[angle=-90,width=7.7cm]{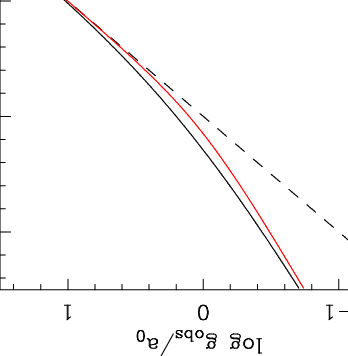}
\caption{Comparison between the RAR (full black curve) and the simple interpolation formula (red curve)
to compute the MOND mass, with $g_{bar}$ on horizontal and $g_{obs}$ on vertical axis. 
Equality between $g_{obs}$ and $g_{bar}$ is represented by the dashed line.} 
\label{fig:mu-plot}
\end{figure}

  In the MOND phenomenology, the Poisson equation $\Delta \phi = 4 \pi G \rho$ is modified in 
\begin{equation}\label{eqn:Poisson}
  \nabla \cdot \left[\mu(\frac{|\nabla\Phi|}{a_0})\nabla \Phi \right] = 4 \pi G \rho , 
  \end{equation}
  with the solution 
  $\mu(\frac{a}{a_0}) \bf{a} = \bf{g_n}$ or $\mu(x)a = g_n$ where $x = a/a_0$.
  The interpolation function $\mu(x)$ tends to x, in the low acceleration regime, and to 1 in the Newtonian limit. 
  To retrieve the MOND acceleration from the Newtonian one, the formula must be inverted to
  $a = \nu(y) g_n$ where $y = g_n/a_0$. This reverse interpolation function
  tends to 1, for $y >> 1$ in the Newtonian regime, and towards $1/\sqrt(y)$ for $y << 1$.
  This solution to the modified Poisson equation is valid modulo a curl function, which
  disappears in spherical symmetry.
  The simple interpolating function is frequently used $\mu(x) = x (1+x^2)^{-1/2}$. This can be inverted 
  in $\nu^2(y) = 0.5 \left[ 1 + \sqrt(1+4/y^2)\right]$. This function is equivalent to the 
  radial acceleration relation (RAR), which is found empirically from the fitting
 of many rotation curves of spiral galaxies. The formula between the observed
 acceleration g$_{obs}$ and the one expected from baryons only in Newtonian dynamics
 g$_{bar}$ takes the form \cite{McGaugh2008ApJ...683..137M}:
 \begin{equation}\label{eqn:RAR}
  g_{obs}= \frac{g_{bar}}{1 - e^{-\sqrt\frac{g_{bar}}{a_0}}}
 \end{equation}
 The two interpolation formulae are compared in Figure \ref{fig:mu-plot}. The simple analytical formula is
 however lower than the RAR, and we have adopted the latter to estimate the required MOND mass. We have 
 numerically inverted, using the Newton-Raphson method, the MOND mass from the dynamical mass.

 With this method, the RAR relation is automatically satisfied for all clusters.

 \subsubsection{The visible mass}
 
 Knowing the MOND mass, it is possible to estimate the "dark" missing mass, by
 subtracting the visible mass, composed mainly of the hot ICM gas, but also in a
 small percentage the mass of the cluster galaxies, and more importantly in the
 center of the Brightest Cluster Galaxy (BCG) combined with the intra-cluster light (ICL).
 
To estimate the mass in stars, we adopted the relation found by
Giodini et al. (2009) \cite{Giodini2009ApJ...703..982G} on a sample of 118 X-ray emitting clusters,
with a mass range of M$_{500}$ between 10$^{13}$ and 10$^{15}$ M$_\odot$.
The stellar mass fraction at R$_{500}$ decreases with increasing total mass as:
		\begin{equation}\label{eqn:fstar}
 	f_\mathrm{500}^\mathrm{stars}=5.0^{+0.1}_{-0.1}\times10^{-2}\,\left(\frac{M_\mathrm{500}}	{5\times10^{13}\,M_{\odot}}\right)^{-0.37\pm0.04} 
	\end{equation}
    
The mass of the BCG combined with the ICL within a radius of 100 kpc has been estimated as a function of M$_{500}$ by
DeMaio et al. (2020) \cite{DeMaio2020MNRAS.491.3751D}, on a sample of 42 groups and clusters:
\begin{equation}\label{eqn:BCG}
\log \textrm{M}_{\star}=0.37\pm0.05\, \log\left( \frac{M_{500}}{2\times10^{14} M_\odot} \right) + 11.92\pm0.02
\end{equation}

Given the mass M inside 100 kpc, it is then distributed along the Hernquist profile M(r) = M r$^2$ /(1 + r$_h$)$^2$, 
with the half-mass radius R$_e$ = (1+$\sqrt{2}$) r$_h$. 

\section{Results and discussion}

The analysis results of the 6 first clusters are presented in Figure \ref{fig:mtot-bar-1}. The whole sample can be found in the Supplementary Information (SI) appendix. The various masses are plotted versus radius, from 20~kpc up to the R$_{500}$ of each cluster. The total mass required under MOND (dashed line) is between 4 and 20 times lower than the one in 
Newtonian dynamics (full black curve). The ratio depends on mass, as can be seen in the top left panel of
Figure \ref{fig:plot-corr}. 
 All results are plotted for the interpolation function corresponding to the RAR. The simple function shown in Fig. \ref{fig:mu-plot} has also been used for the 106 clusters, and gives similar results, with slightly 
more dark missing mass. The ratio M$_\text{{dark}}$/M$_\text{{MOND}}$ is on average 0.53, compared to 0.41 for
the RAR function.

 From the ratio between these two values, bracketing the possible interpolation functions, we derive
an error bar for the M$_\text{MOND}$ determination of 0.11 dex. For M$_{500}$, M$_\text{vis}$ and $\sigma_v$, 
the error bars are the average of the errors given in \cite{Chen2007A&A...466..805C}.
Then we estimated at 0.22 dex the errors in deriving M$_\text{dark}$, from M$_\text{dyn}$-M$_\text{MOND}$ at each radius,

\subsection{Mass inconsistencies and dark mass profile}
At R$_{500}$, the visible mass including ICM, galaxies and BCG+ICL (red curve) continues to increase and, in some cases, exceeds the required MOND mass towards the outer radii, thereby invalidating the hypothesis. This discrepancy occurs in the cluster periphery, where the gas mass remains uncertain and its assumed hydrostatic equilibrium is questionable. Similarly, in the cluster core, particularly for non-CC clusters with large core radii, the visible mass surpasses the dynamical mass. Within this central region, the visible mass is dominated by the BCG; consequently, the scaling relation used (\ref{eqn:BCG}) may introduce significant scatter. Nevertheless, this central inconsistency has a negligible effect on the overall mass budget, as the BCG accounts for less than 1\% of the total mass. 

The difference between the required MOND mass, M$_{\text{MOND}}(r)$, and the visible mass, M$_{\text{vis}}$, is calculated radially as $\delta M(r)$ (green curve) up to the turning point, where this accumulated mass reaches a plateau. The density profile of this missing mass, $\rho_{\text{dark}}$, is given by its derivative: $\rho_{\text{dark}}(r) = \frac{d\delta M(r)}{4\pi r^2 dr}$. Beyond the turning point, the underlying hypothesis breaks down, causing the MOND “phantom” mass density to become negative. The resulting missing mass density profile (blue curve) exhibits a central core with a steep truncation at the ICM core radius, well within R$_{500}$, aligning with previous findings.

\begin{figure*}
\centering
\includegraphics[width=16.0cm,angle=0]{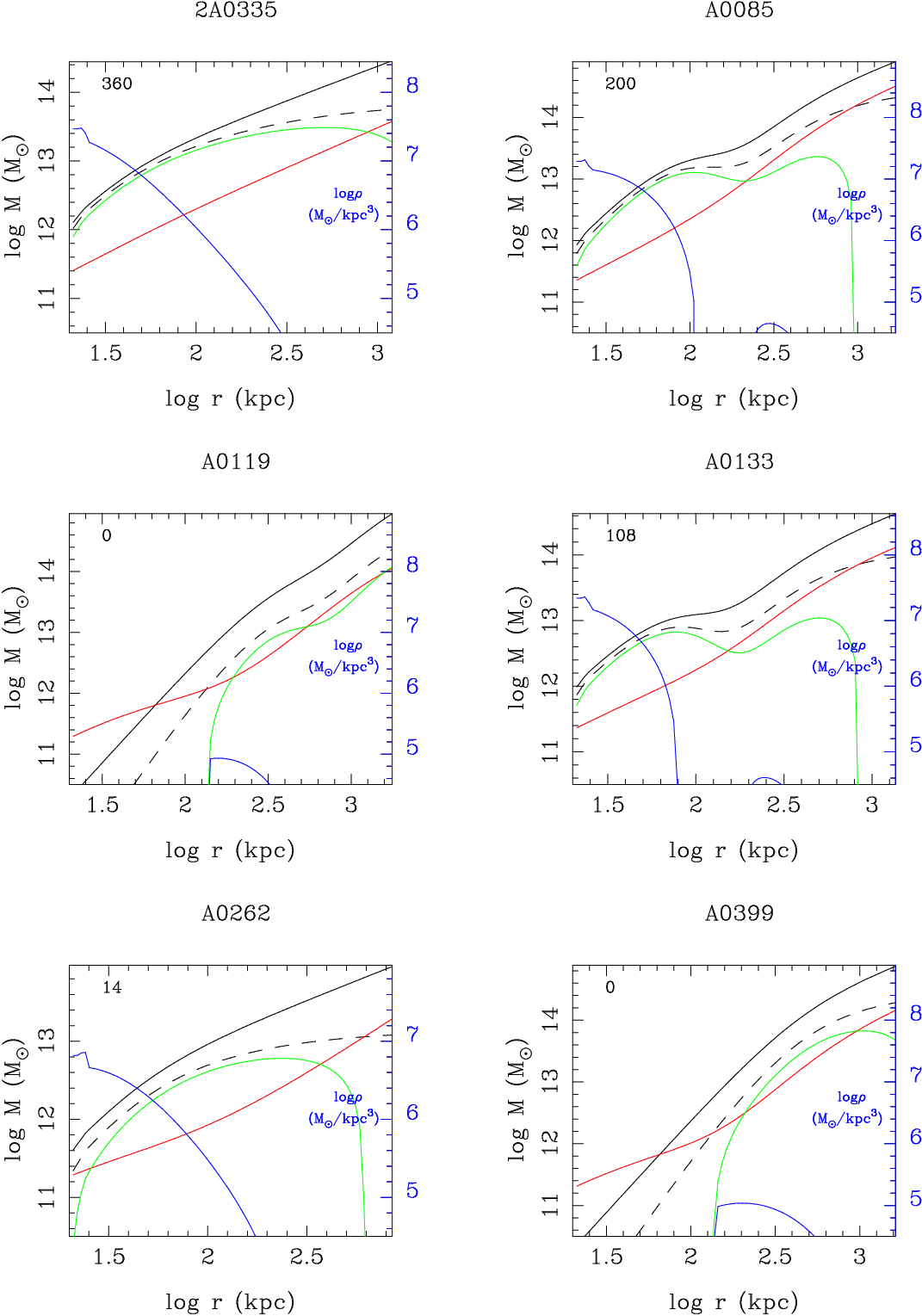}
\caption{Results for the 6 first clusters. The radius begins at 20~kpc, and ends at R$_{500}$. The full black curve is the Newton dynamical mass M(r) derived assuming hydrostatic equilibrium,  including both the temperature gradient and the second beta-model component, when it improves the fit with respect to only one component. The dashed line is the mass M$_\text{{MOND}}$(r) necessary in MOND to account for this equilibrium, following equation \ref{eqn:RAR}. The red curve is the visible mass M$_\text{{vis}}$ including the X-ray emitting hot gas, the galaxies of the clusters and the BCG+ICL. The green curve is then the difference between M$_\text{{MOND}}$(r) and M$_{\text{vis}}$ and represents the "dark" mass. By derivation at each radius, the dark density is obtained, and plotted in blue, with the scale to be read at right. The value of $\dot{M}$ is indicated at the top left of each panel. }
\label{fig:mtot-bar-1}
\end{figure*}

Within the extended HIFLUGCS sample, targets are divided almost equally between non-CC systems and those exhibiting weak or strong cool cores (Table \ref{Tab:CC}). Our computations reveal distinct discrepancy patterns across these categories: every non-cool-core cluster exhibits a mass inconsistency in its central region, with a subset showing additional discrepancies in the outskirts. Conversely, cool-core clusters remain predominantly consistent, allowing for a robust determination of the dark mass density profile throughout their entire volume.

\begin{table}[t!] 
\centering
\caption{Comparison of the cool-core and non-cool core clusters} \label{Tab:CC}
\begin{tabular}{lrrr}
       & NCC & weak-CC & CC \\
\midrule
Number of clusters                             & 56 & 25 & 25 \\
M$_\text{{vis}}$ lower than M$_\text{{MOND}}$  & 0  & 8  & 14 \\
BCG+ICL overmassive                            & 56 & 12 & 8  \\
Outer parts overmassive                        & 8  & 7  & 6  \\
\bottomrule
\end{tabular}

\addtabletext{$\dot{M}$=0 (NCC), $<$ 60 (weak-CC) and $>$ 60 M$_\odot$ yr$^{-1}$ (CC) }
\end{table}

\begin{figure}
\centering
\includegraphics[width=8.7cm,angle=0]{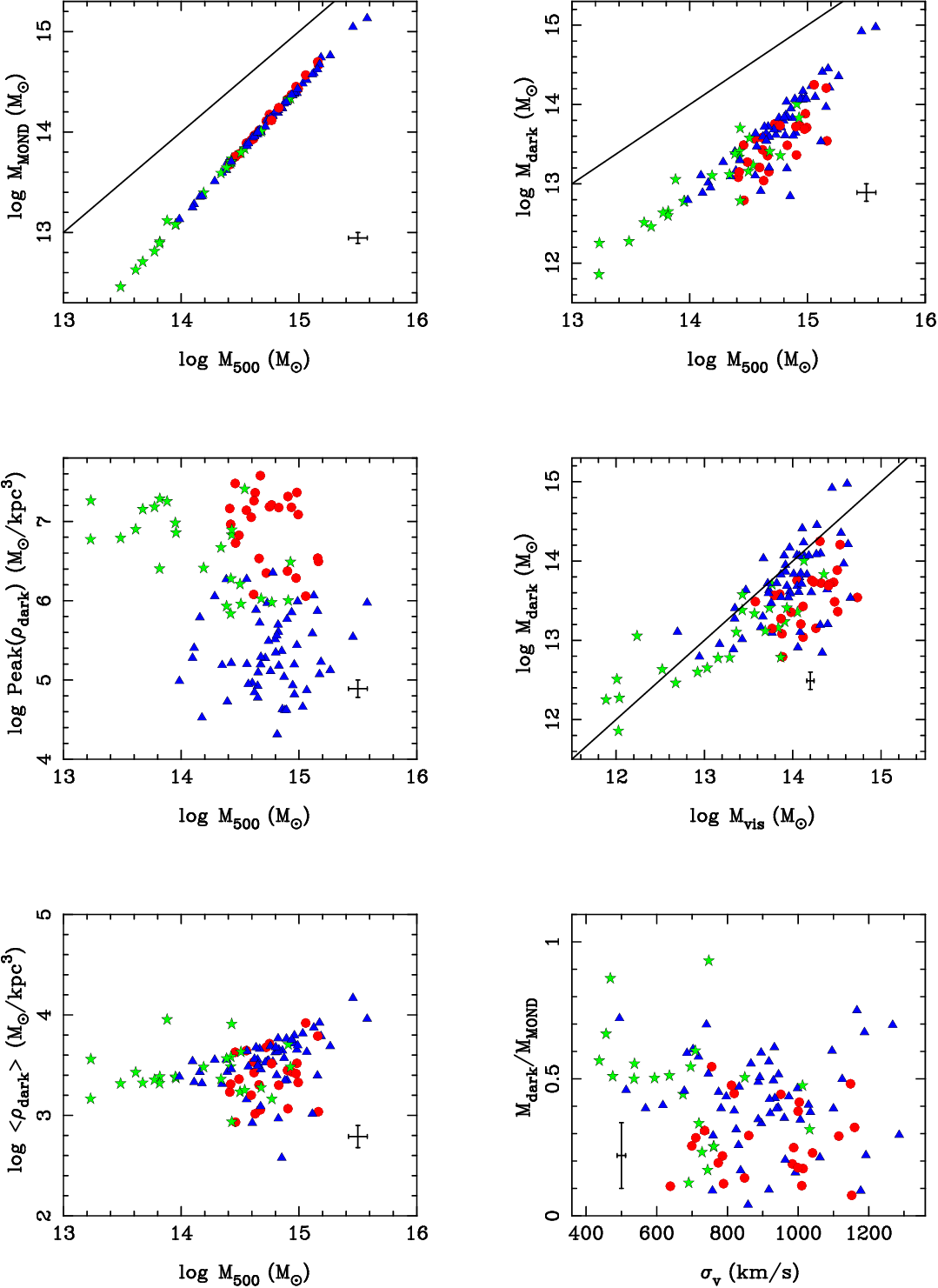}
\caption{Correlations of the total mass computed with MOND M$_\text{{MOND}}$ and the derived dark mass versus the dynamical mass M$_{500}$ computed with Newtonian dynamics within the R$_{500}$ radius, and versus the visible mass M$_\text{{vis}}$ at R$_{500}$. Also plotted are the derived peak and average density of dark mass,
and the ratio M$_\text{{dark}}$/M$_\text{{MOND}}$ as the function of cluster velocity dispersion $\sigma_v$.
The non-CC are blue triangles, the weak-CC green stars, and the CC red bullets.} 
\label{fig:plot-corr}
\end{figure}

These three sub-populations are color-coded and assigned distinct symbols in Figure \ref{fig:plot-corr}. As illustrated in the middle-left panel, the dark mass density peaks at noticeably higher levels in cool-core clusters, with strong cool cores preferentially residing in the most massive systems. The top-right panel demonstrates that while dark mass scales with total cluster mass, it does so with a super-linear slope—yielding a slightly higher relative fraction of dark mass in high-mass clusters, where global acceleration exceeds the critical threshold $a_0$. The middle-right panel highlights the relationship between dark and visible mass; the dark mass closely tracks the visible component, generally maintaining a magnitude roughly twice lower.
The average
dark mass density (bottom-left) also increases with total mass, and the M$_\text{{dark}}$/M$_\text{{MOND}}$
(bottom-right) does not depend on the cluster velocity dispersion. Since the latter is still non-relativistic,
it is not possible to gain a clue towards the relativistic MOND behavior.

\subsection{Caveats}

Uncertainties on the derived total dynamical mass, and therefore MOND mass and dark mass,
might be larger than the error bars derived for a given method, due to adopted simplifications:
hydrostatic equilibrium, spherical symmetry, homogeneity of the hot gas.
For instance, a recent computation of the dynamical mass with the galaxy kinematics method
for a few HIFLUGCS clusters has found up to 50-80\% differences with the hydrostatic equilibrium
method \cite{Li2023A&A...677A..24L}. In general, the kinematic method gives larger masses, possibly
due to the galaxy ensemble being out of equilibrium, although the hot ICM appears in 
hydrostatic equilibrium. It is possible that the various components do not have the same time-scale
to reach equilibrium.
 There are uncertainties in the determination of temperature and its gradients due to turbulence
\cite{Lovisari2024A&A...682A..45L}. 

 The visible mass, mainly the hot ICM mass, is relatively well determined, and the inhomogeneities
are not significant \cite{Eckert2015MNRAS.447.2198E}. However, the total
gas mass comes essentially from the outer parts, where the gas becomes diffuse and is more difficult to detect.
The density slope is still not steep enough at R$_{500}$ to prevent the total mass from being dominated
by these regions, which are variable from cluster to cluster \cite{Ghirardini2019A&A...621A..41G}.
The outer parts are subject to accretion from cosmic filaments, shock heating, when
not merging with sub-clusters. The majority clusters reveal sloshing perturbations, even the relaxed
ones \cite{Ueda2020ApJ...892..100U}.

Even within the Newtonian dynamics with dark matter, the baryonic fraction is exceeding the cosmic
universal value because of an excess of hot gas in the cluster periphery. This is interpreted
as evidence for the contribution of nonthermal pressure at the level of 10-20\% \cite{Eckert2022A&A...662A.123E}.

The dynamical anomalies observed in the outer regions of the cluster may find their resolution through the External Field Effect (EFE)—a prediction of MOND arising from its violation of the Strong Equivalence Principle.
Unlike General Relativity, MOND’s non-linear nature causes a system's internal gravitational dynamics to depend on the surrounding cosmic environment. Thus, while the Weak Equivalence Principle holds, an object's internal motion remains bound to its specific spatial location. Although calculating the EFE in three dimensions generally demands complex numerical simulations, one-dimensional approximations offer a reliable order of magnitude \cite{Famaey2012LRR....15...10F}.

Crucially, the EFE alters the galaxy's potential at large radii, transitioning it from logarithmic behavior back to a Newtonian falloff, and thereby defining a clear MOND escape velocity. In the outskirts of galaxy clusters, this effect may ultimately eliminate the requirement for negative "phantom" mass \cite{Kelleher2024A&A...688A..78K}.  Because of the large number of unknowns and the fact that spherical symmetry is then broken, the estimation of this effect has to wait for detailed numerical simulations, beyond the scope of this work.

\section{Candidates for the dark baryons}

Although the hot intracluster medium (ICM) constitutes the vast majority of visible mass in rich galaxy clusters, exceeding galactic mass by a factor 10, a substantial reservoir of dark baryonic matter may still remain hidden. To reconcile the observed dynamics of these clusters with theories such as MOND, three primary reservoirs for this missing mass have been proposed.

(1)	Stellar Remnants: Applying the integrated galaxy-wide initial mass function (IGIMF), Zhang et al. (2026) \cite{Zhang2026PhRvD.113d3027Z} suggest that remnants from galaxies and intracluster light could account for a dark baryon mass equivalent to 60\% of the ICM. This would simultaneously alleviate the missing mass deficit and naturally explain the extreme metallicity observed within the ICM.

(2)	Ultra-Diffuse Galaxies: The recent discovery of thousands of highly dispersed, ultra-diffuse galaxies within clusters offers another potential harbor for these hidden baryons \cite{Milgrom2015MNRAS.454.3810M}. 

Let us note that standard spiral galaxies can host no more than 1–2\% of these missing baryons; exceeding this threshold would over-predict galactic rotation speeds, violating the Tully-Fisher relation, which requires the total gas surface density to remain roughly three times that of atomic gas alone \cite{Tiret2009A&A...496..659T}.

(3)	Cold Molecular Clouds: A substantial fraction of the missing mass could exist as tiny dense and cold gas clouds, multi-phase, with the bulk of the mass in molecular form. Organized in a hierarchical fractal structure with an extremely low volume-filling factor, this multi-phase gas would behave in an almost entirely dissipationless manner \cite{Pfenniger1994A&A...285...79P, Pfenniger1994A&A...285...94P}.

A fraction of this cold gas is already detectable in cool-core clusters. In these environments, cooling hot gas flows toward the central Brightest Cluster Galaxy (BCG) to feed its supermassive black hole, while subsequent active galactic nucleus (AGN) feedback violently expels metal-rich gas across 100-kpc scales \cite{Fabian2012ARA&A..50..455F}. This cyclical engine sustains an enduring reservoir of cold, enriched molecular gas. In the prototypical Perseus cluster, $10^{10}$ to $10^{11}$ M$_\odot$ of molecular gas has already been observed via carbon monoxide emission at 10-kpc scales \cite{Salome2006A&A...454..437S, Russell2019MNRAS.490.3025R, Olivares2019A&A...631A..22O}.

Extrapolating this mechanism to fully resolve the cluster mass discrepancy requires roughly $10^{14} M_\odot$ of cold molecular gas within a 500-kpc radius. This translates to an ensemble of $10^{17}$ individual clouds, each bearing a Jupiter mass ($10^{-3}$ M$_\odot$) and a radius of just a few astronomical units, at temperatures $\sim$ 15 K. Despite their vast numbers, their combined sky-covering factor in the cluster core would remain about $10^{-4}$.

Ultimately, high-resolution numerical simulations are required to determine whether these highly concentrated dark baryons can fully reconcile MOND with cluster dynamics, properly accounting for the system's broken symmetries, inherent inhomogeneities, and out-of-equilibrium peripheries.



\section{Summary}

Using the extended HIFLUGCS sample of 106 galaxy clusters, we have derived the residual missing mass within the framework of MOND. Assuming spherical symmetry and hydrostatic equilibrium for the hot ICM medium, we first established total cluster masses under Newtonian gravity. The MOND-required dynamical mass was subsequently computed via the modified Poisson equation, anchored by two physical boundary conditions: at accelerations well above the critical threshold ($a_0 = 1.2 \times 10^{-10} \text{m/s}^2$), the gravitational field converges to the Newtonian value ($g_n$); in the deep MOND regime ($a \ll a_0$), it approaches $a = \sqrt{a_0 g_n}$. To bridge these regimes, we adopted the Radial Acceleration Relation (RAR) interpolation function, an empirical law proven to hold across spiral galaxies.

The residual missing mass represents the deficit between the MOND-required mass and the observed baryonic mass—the latter comprising the dominant hot ICM, member galaxy masses, and the central Brightest Cluster Galaxy (BCG) alongside the intracluster light (ICL). This required dark component is sharply concentrated toward the cluster center, exhibiting a steep drop-off near the ICM core radius (a few tens of kiloparsecs), well inside $R_{500}$. Modified gravity reduces the required dynamical mass by a factor of 4 to 10, the reduction being greatest for the least massive clusters. A substantial dark mass residual remains, proportional to the total gas mass and about a factor of two lower on average.

Our sample further reveals distinct behavior between cluster morphology types. Cool-Core (CC) clusters display the steepest central dark mass density peaks and yield the most robust fits under our reconstruction procedure. Non-Cool-Core (non-CC) clusters possess larger ICM core radii. Here, the central BCG mass dominates to such an extent that it exceeds both the MOND requirement and the Newtonian dynamical mass. This unphysical localized overshoot likely stems from systematic uncertainties in modeling BCG+ICL masses via scaling relations, rather than a global mass budget failure. At the cluster periphery ($\sim$ R$_{500}$), the observed hot gas mass continues to rise until it surpasses the MOND limit. In these outer regions, the validity of hydrostatic equilibrium and standard gas density profiles becomes increasingly questionable, rendering the derivation unreliable. In these outskirts, the External Field Effect should change the dynamics, and might solve the anomalies. 

To reconcile the residual mass without invoking collisionless dark matter, several hypotheses propose dark baryons; the most plausible would be cold, dense molecular gas clouds that behave effectively like a dissipationless fluid. Such a component would harmonize cluster dynamics with a unified empirical scaling relation, smoothly extending the Baryonic Tully-Fisher Relation from galactic scales up to galaxy clusters.


\acknow{I am grateful to the two referees for constructive comments.
This work has been inspired by discussions during the Lorentz Center Conference in Leiden "MOND: alternative paths in the dark matter problem".}

\showacknow{} 


\bibliography{pnas-sample}

\section*{Complementary Figures}
 Results for the 6 first clusters were displayed in the main article. Here are displayed the remaining 100 clusters,
 in 17 figures. The name of the clusters are indicated on each figure caption, for easy retrieval.\\

The common caption is \\

Results of the computation for the 100 remaining clusters. The abcissa is the log of the radius beginning with 20~kpc, and ending at the R$_{500}$ given in Chen et al. (2007). The full black curve is the total dynamical mass M(r) derived with Newtonian dynamics, assuming hydrostatic equilibrium, including both the temperature gradient and the second beta-model component, when it improves the fit with respect to only one component. The dashed line is the mass M$_\text{{MOND}}$(r) necessary in MOND to account for this equilibrium, using the RAR as interpolation function. The red curve is the visible mass M$_\text{{vis}}$ including the X-ray emitting hot gas, the galaxies of the clusters and the BCG+ICL. The green curve is then the difference between M$_\text{{MOND}}$(r) and M$_{\text{vis}}$ and represents the "dark" mass. By derivation at each radius, the dark density is obtained, and is plotted in blue, with the scale to be read at right. The value of $\dot{M}$ is indicated at the top left of each panel.
 
\begin{figure*}
\centering
\includegraphics[width=16.5cm]{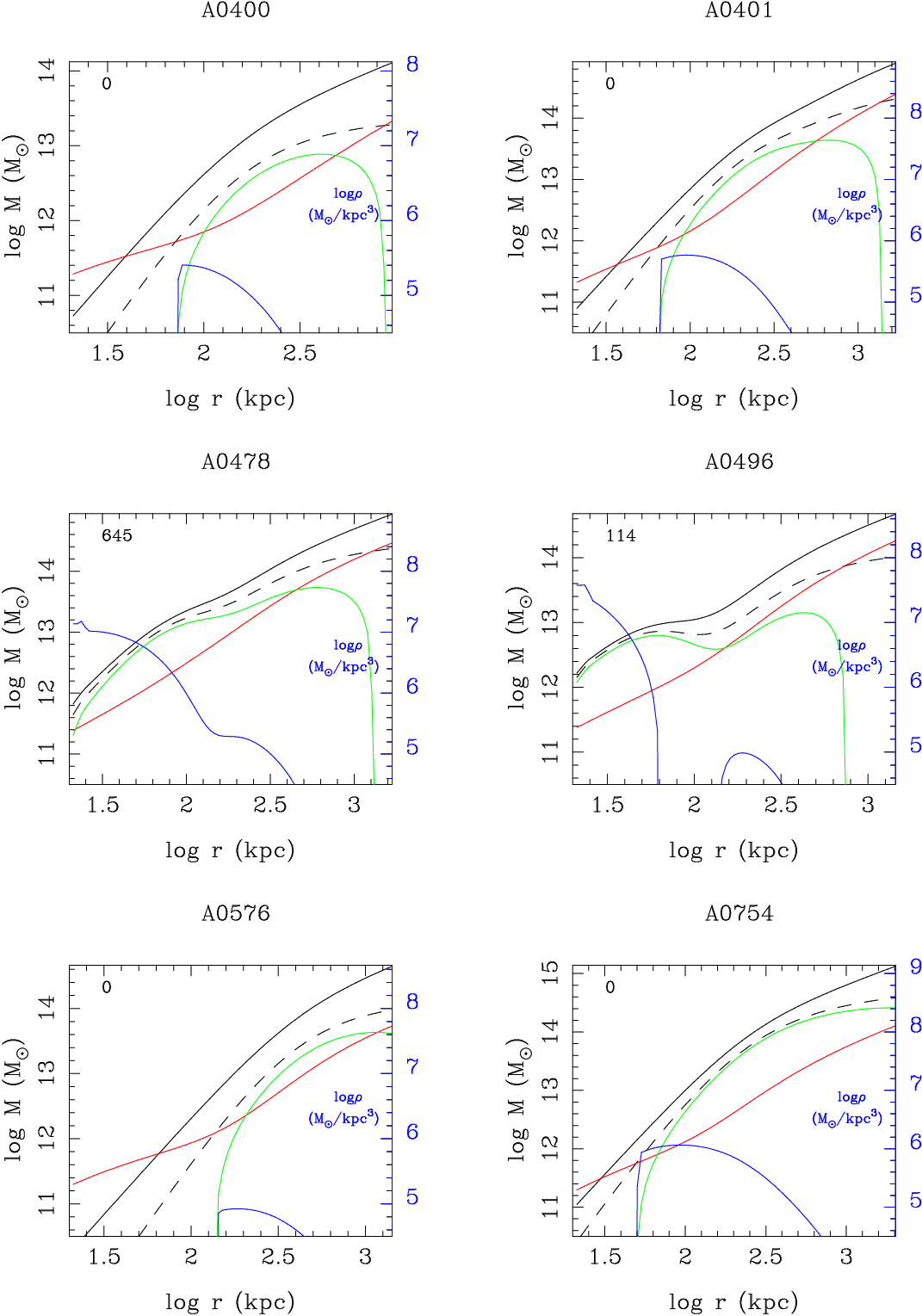}
\caption{A0400   , A0401   , A0478   , A0496   , A0576   , A0754 }
\end{figure*}

\begin{figure*}
\centering
\includegraphics[width=16.5cm]{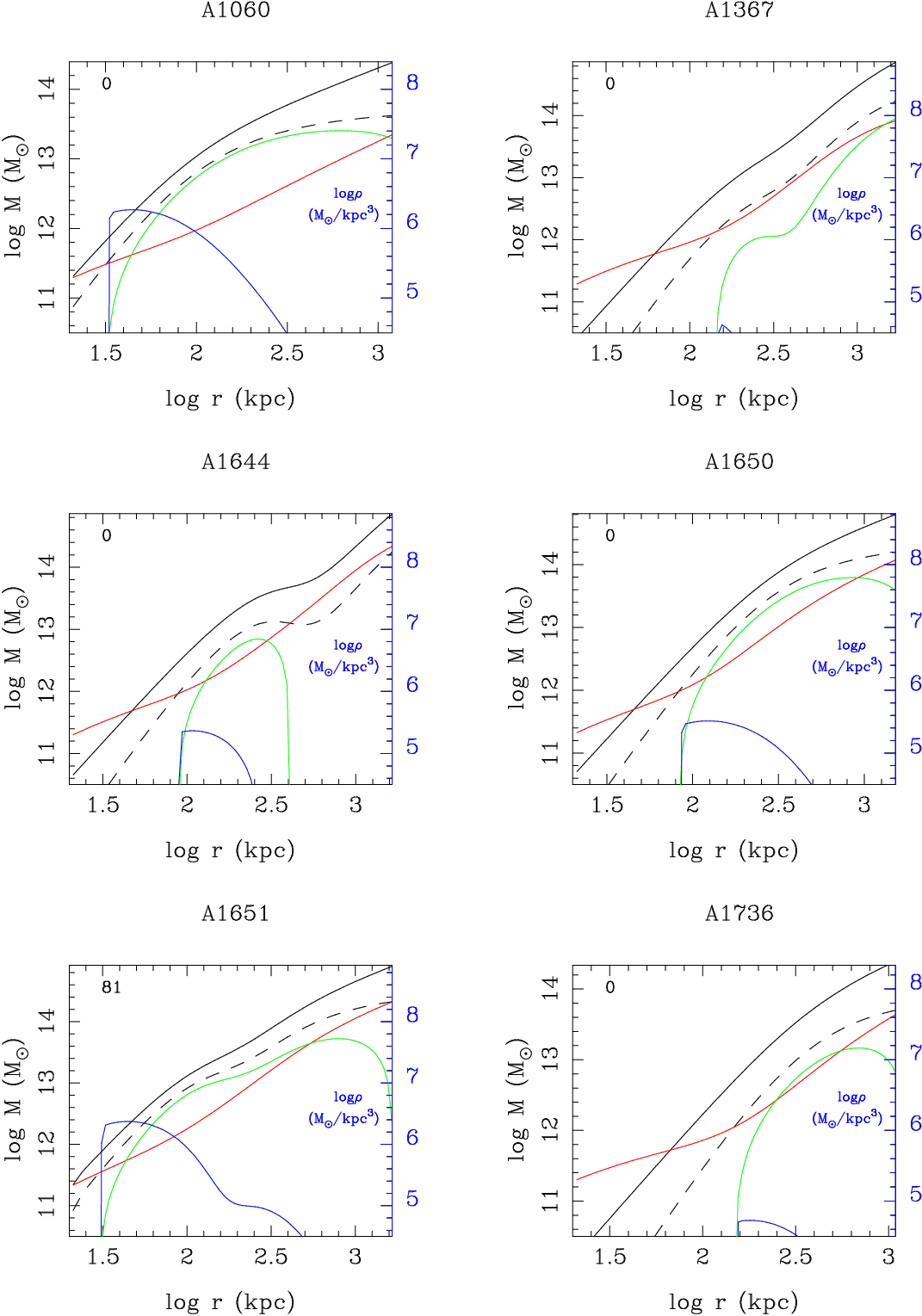}
\caption{ A1060   , A1367   , A1644   , A1650   , A1651   , A1736}
\end{figure*}

\begin{figure*}
\centering
\includegraphics[width=16.5cm]{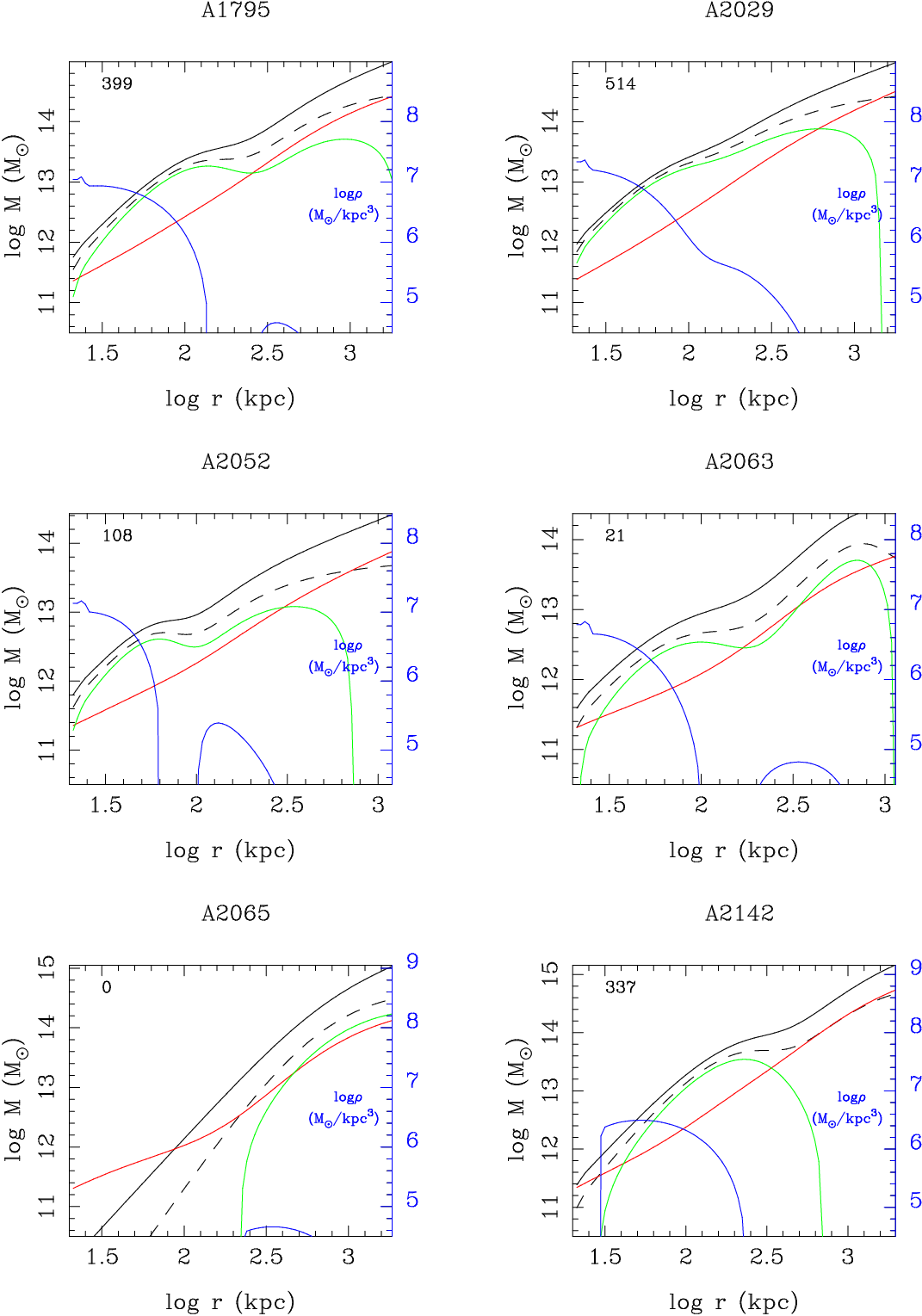}
\caption{A1795   , A2029   , A2052   , A2063   , A2065   , A2142 }
\end{figure*}

\begin{figure*}
\centering
\includegraphics[width=16.5cm]{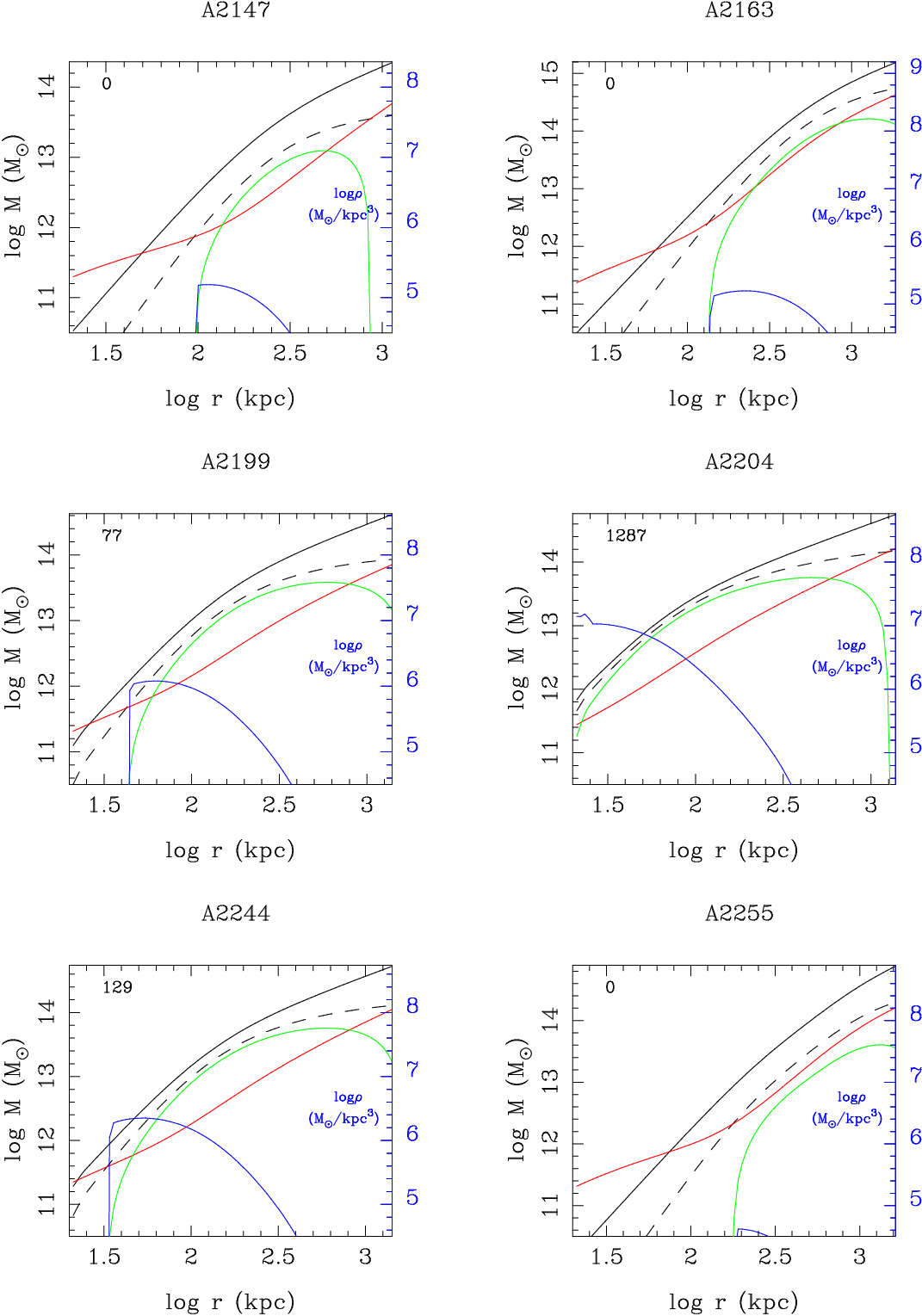}
\caption{ A2147   , A2163   , A2199   , A2204   , A2244   , A2255}
\end{figure*}

\begin{figure*}
\centering
\includegraphics[width=16.5cm]{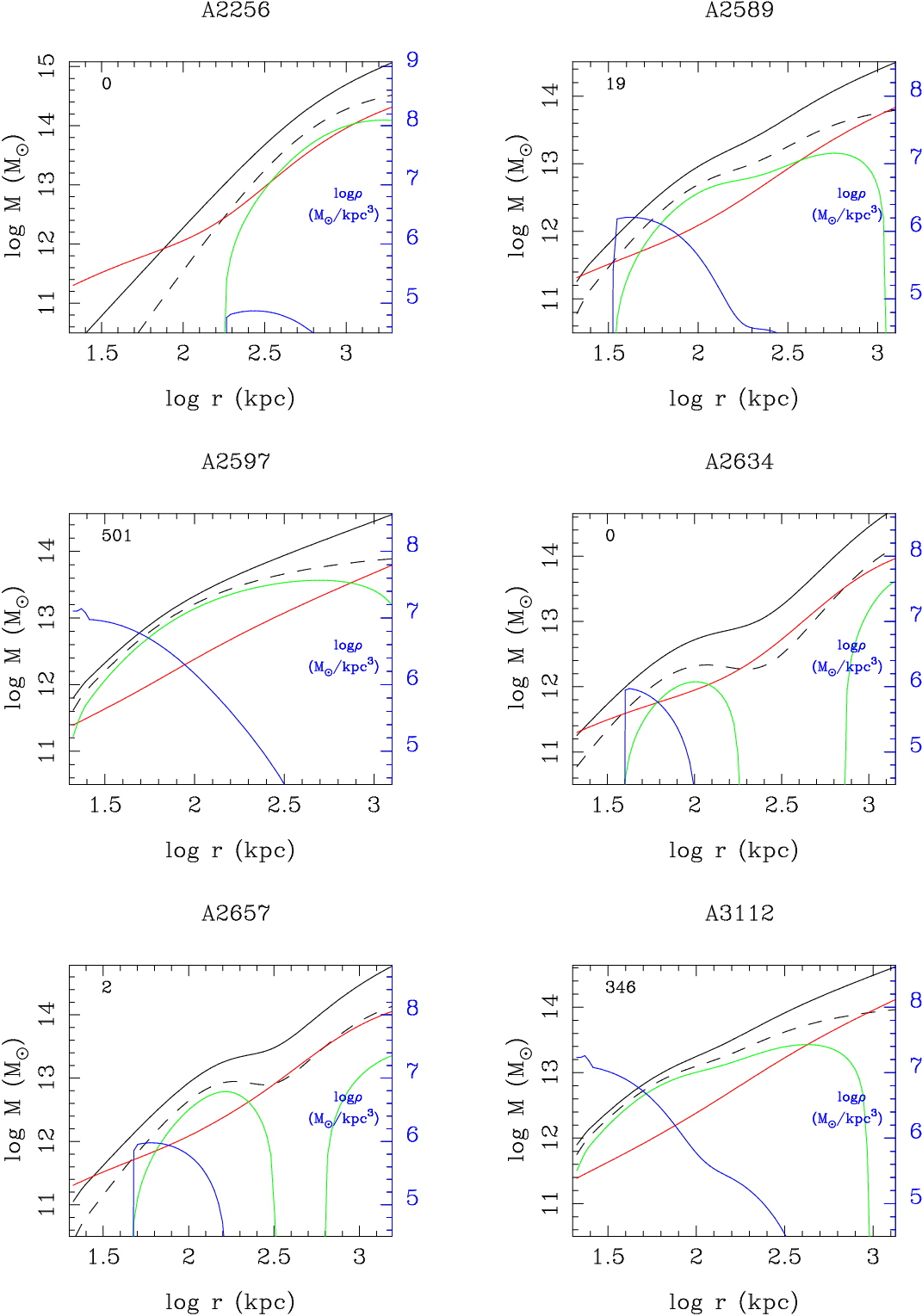}
\caption{A2256   , A2589   , A2597   , A2634   , A2657   , A3112 }
\end{figure*}

\begin{figure*}
\centering
\includegraphics[width=16.5cm]{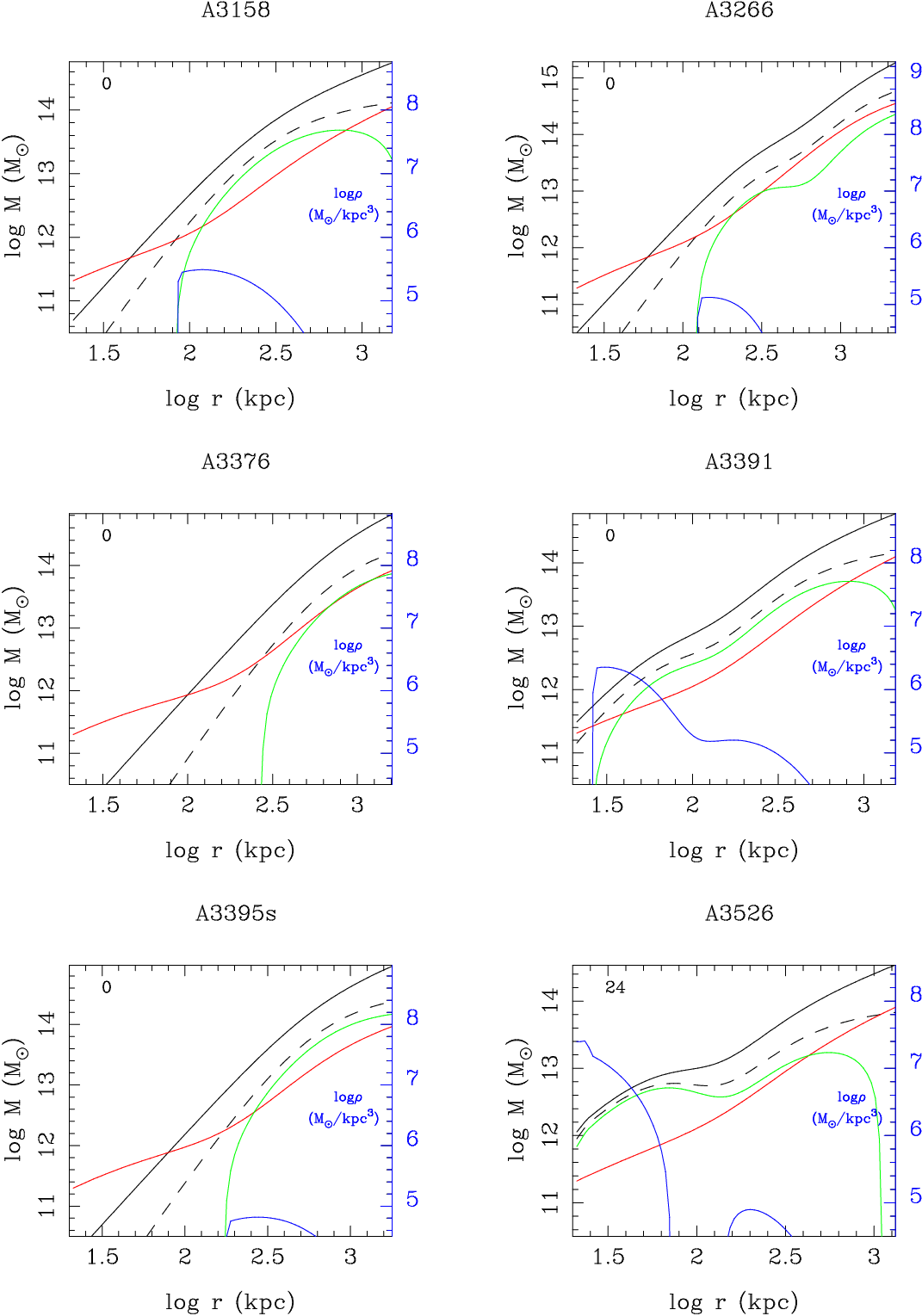}
\caption{ A3158   , A3266   , A3376   , A3391   , A3395s  , A3526 }
\end{figure*}

\begin{figure*}
\centering
\includegraphics[width=16.5cm]{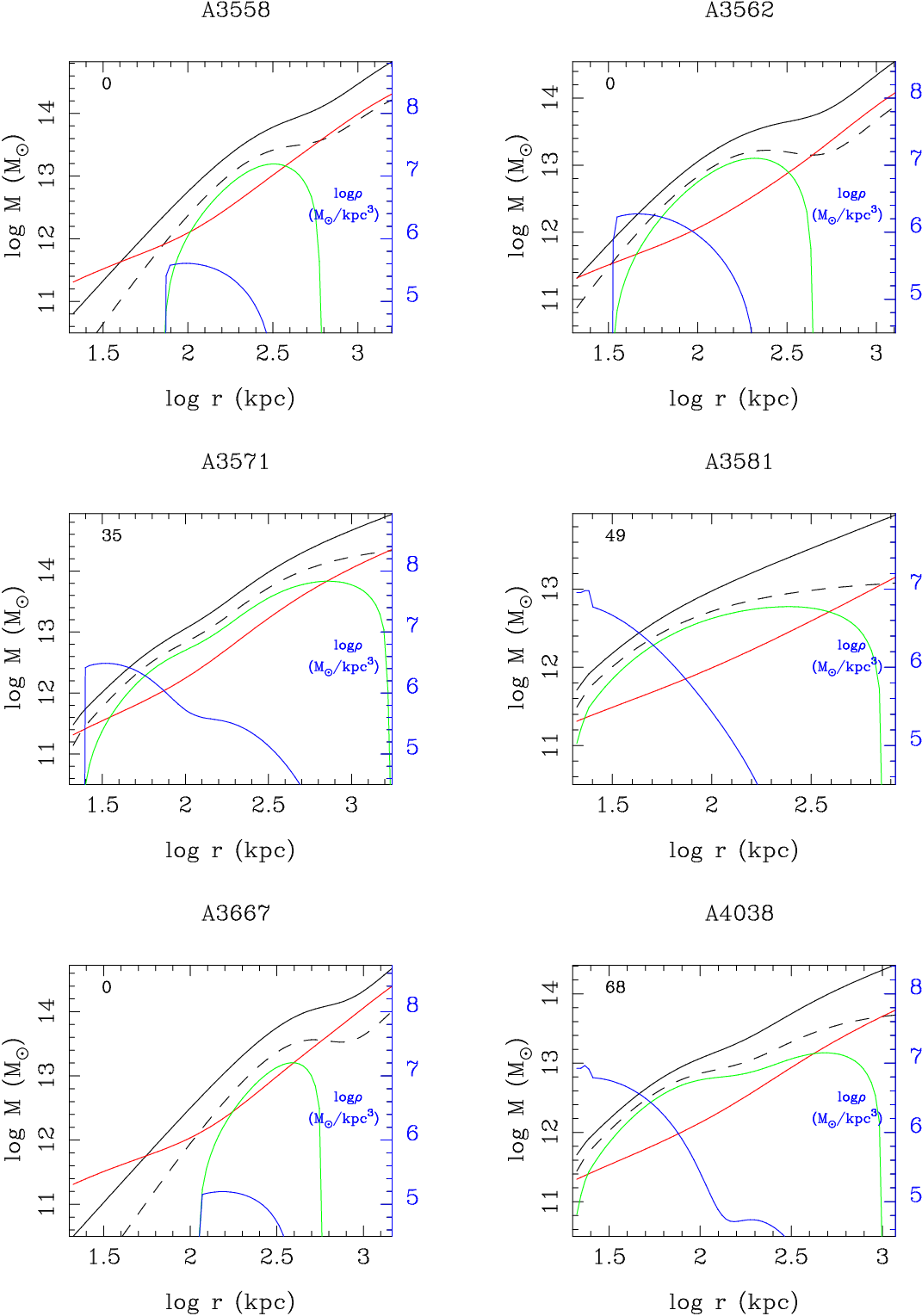}
\caption{A3558   , A3562   , A3571   , A3581   , A3667   , A4038}
\end{figure*}

\begin{figure*}
\centering
\includegraphics[width=16.5cm]{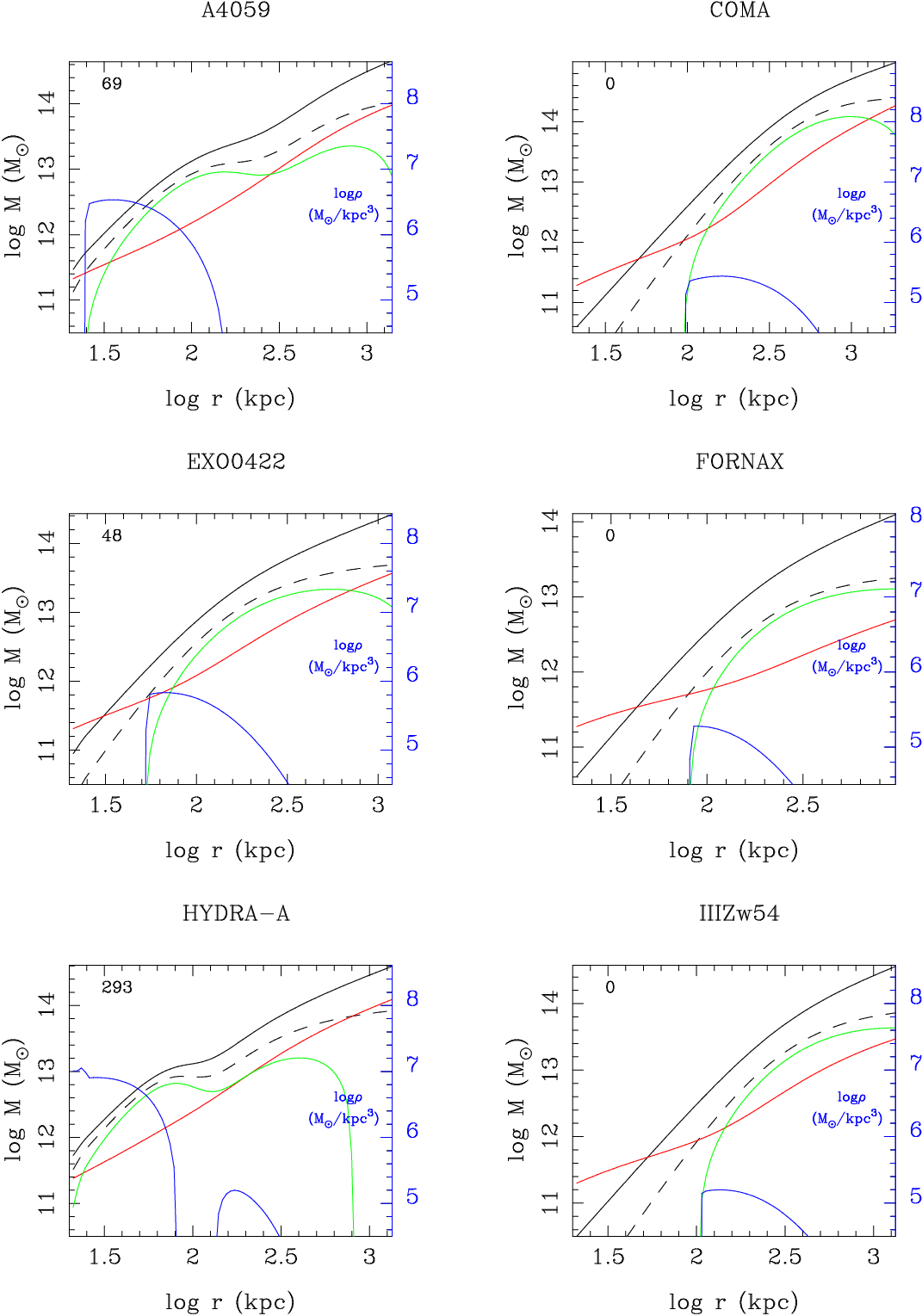}
\caption{A4059   , COMA    , EXO0422 , FORNAX  , HYDRA-A , IIIZw54}
\end{figure*}

\begin{figure*}
\centering
\includegraphics[width=16.5cm]{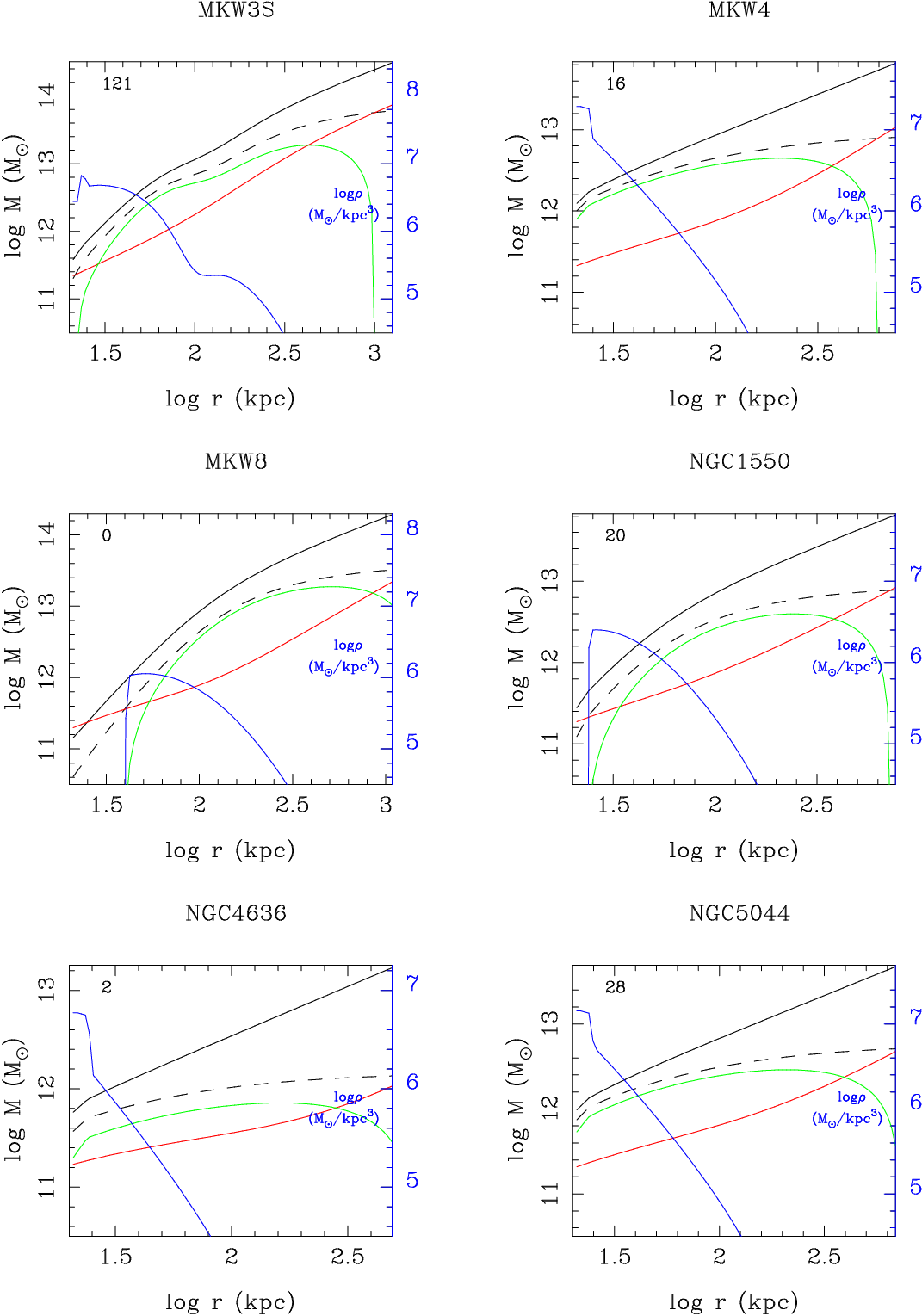}
\caption{ MKW3S   , MKW4    , MKW8    , NGC1550 , NGC4636 , NGC5044}
\end{figure*}

\begin{figure*}
\centering
\includegraphics[width=16.5cm]{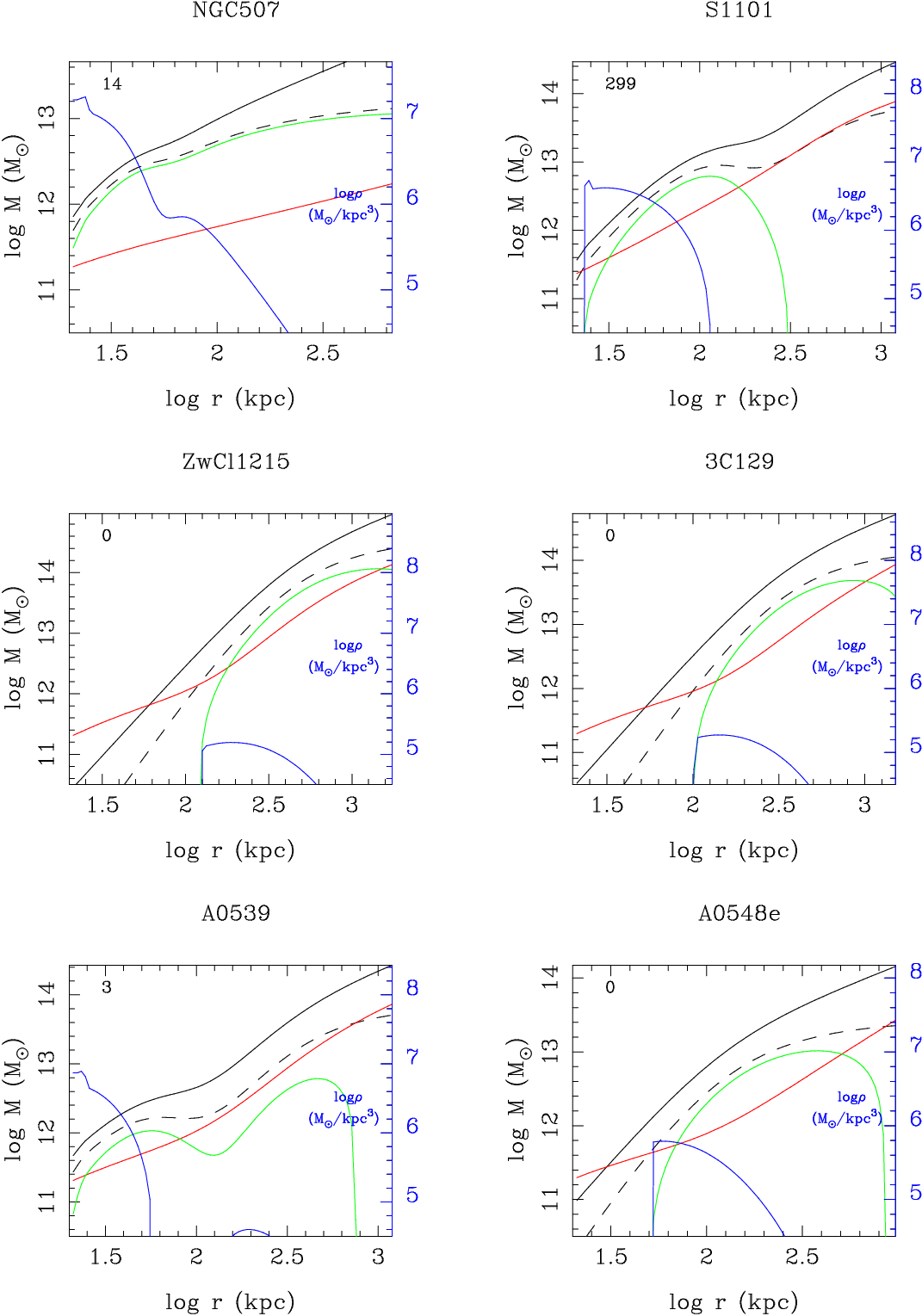}
\caption{NGC507  , S1101   , ZwCl1215, 3C129   , A0539   , A0548e}
\end{figure*}

\begin{figure*}
\centering
\includegraphics[width=16.5cm]{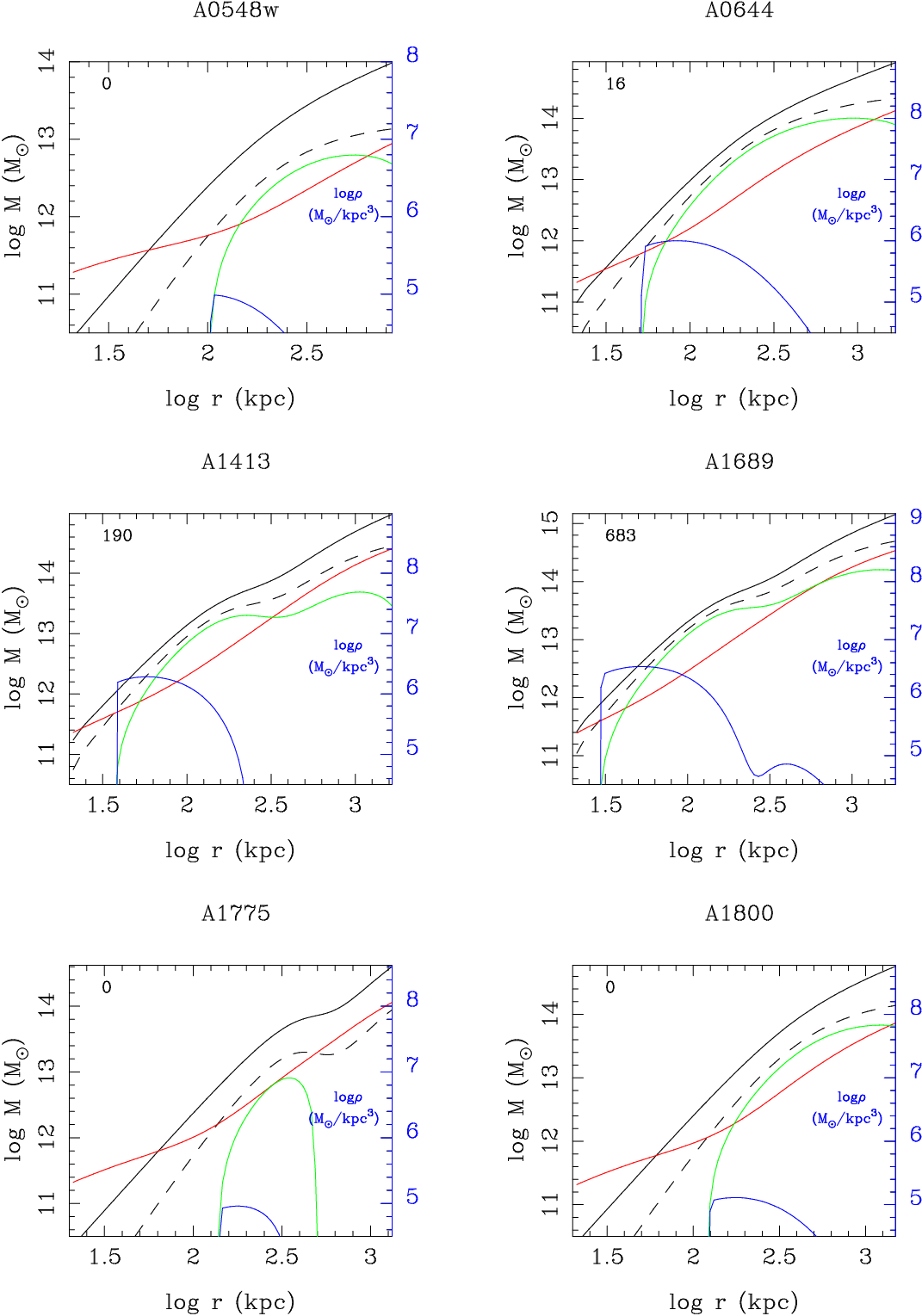}
\caption{A0548w  , A0644   , A1413   , A1689   , A1775   , A1800 }
\end{figure*}

\begin{figure*}
\centering
\includegraphics[width=16.5cm]{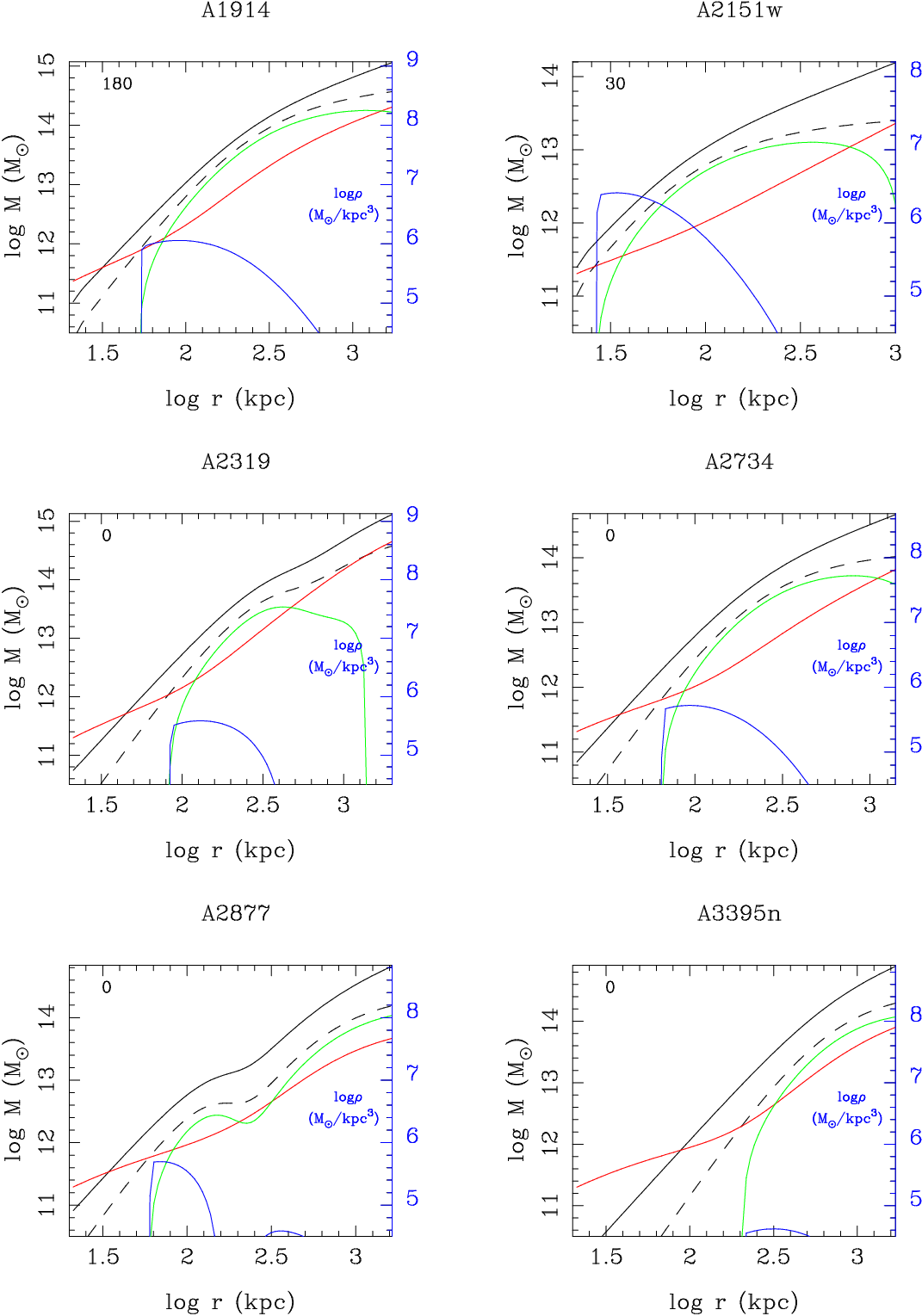}
\caption{A1914   , A2151w  , A2319   , A2734   , A2877   , A3395n }
\end{figure*}

\begin{figure*}
\centering
\includegraphics[width=16.5cm]{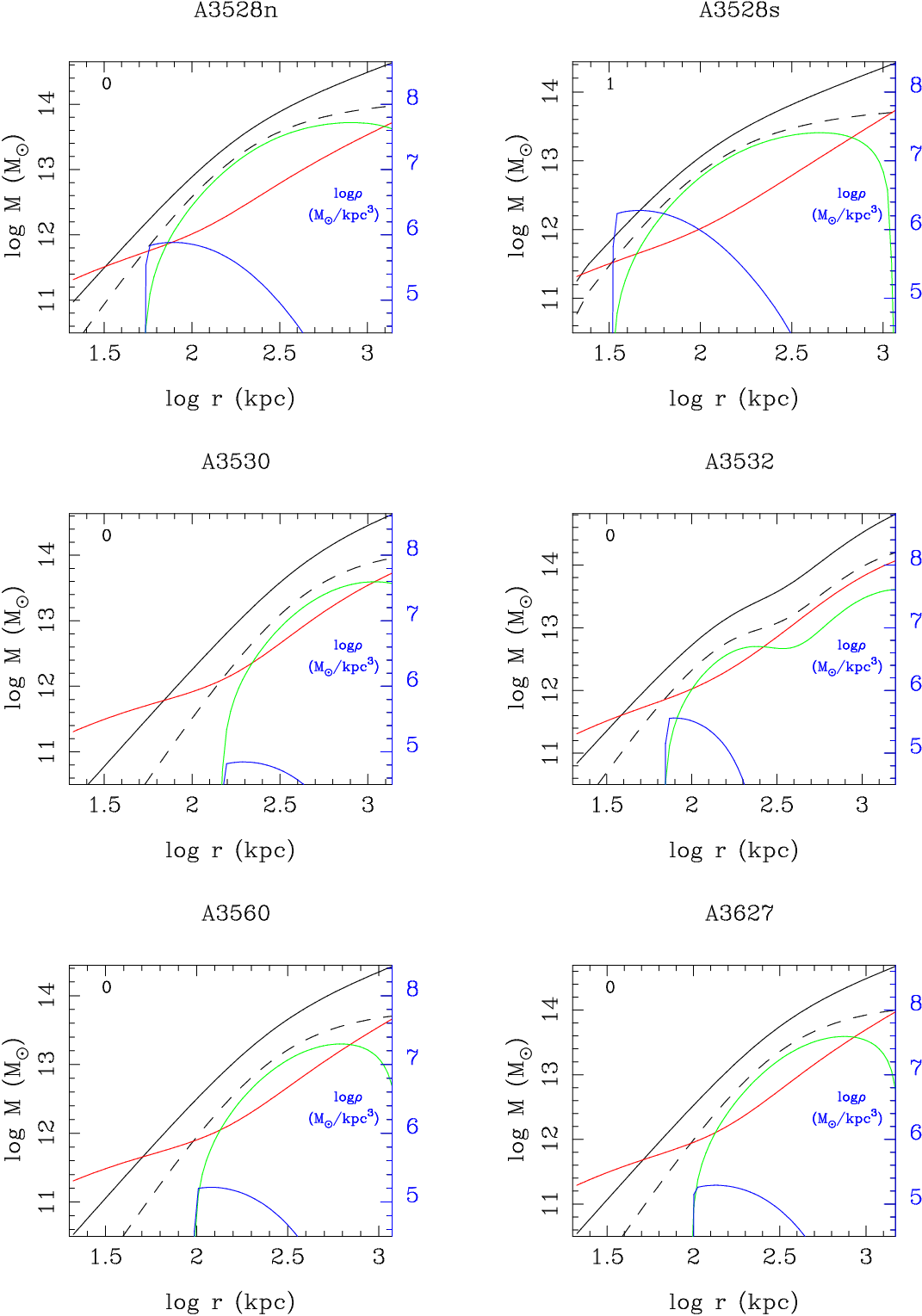}
\caption{ A3528n  , A3528s  , A3530   , A3532   , A3560   , A3627}
\end{figure*}

\begin{figure*}
\centering
\includegraphics[width=16.5cm]{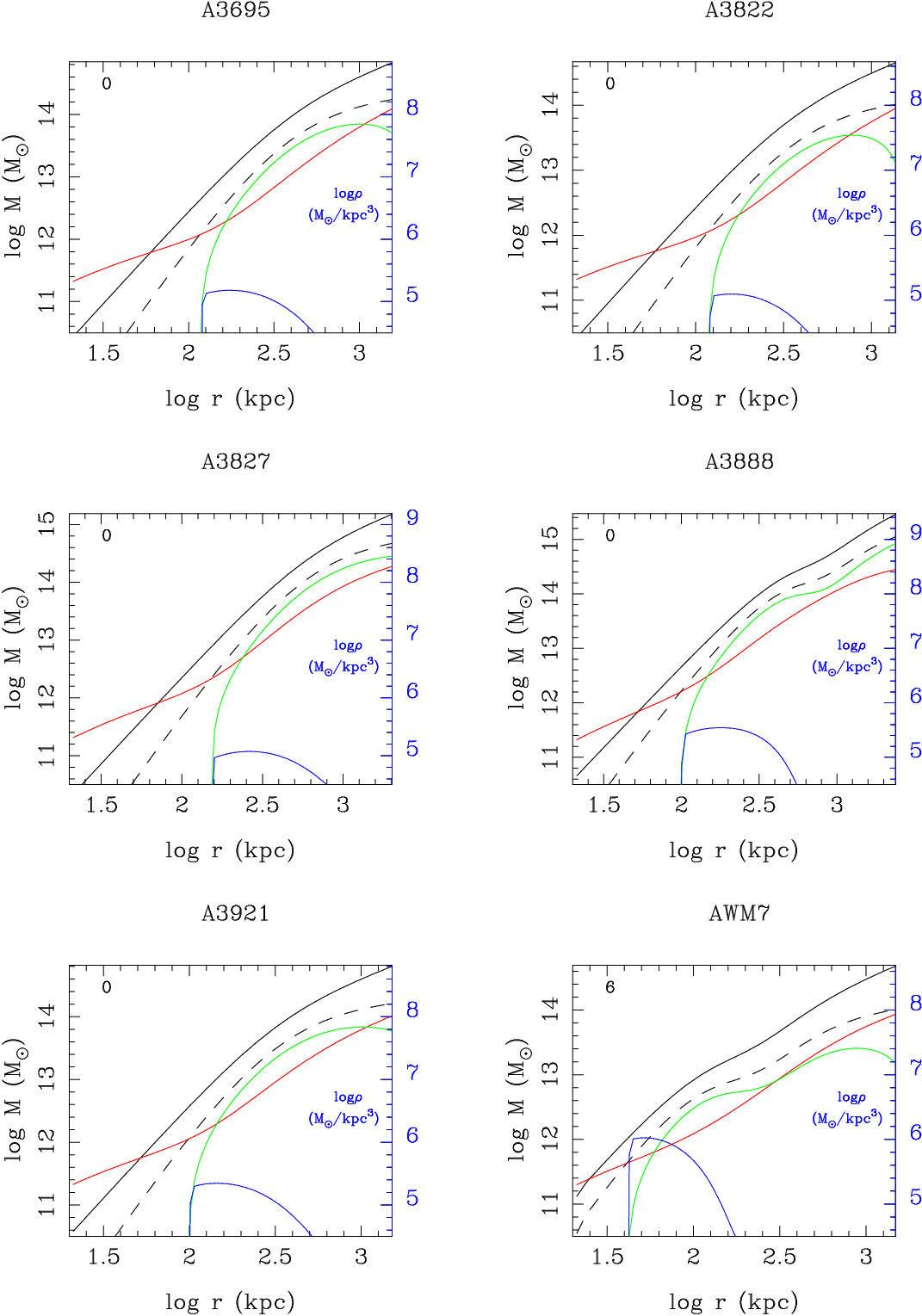}
\caption{A3695   , A3822   , A3827   , A3888   , A3921   , AWM7}
\end{figure*}

\begin{figure*}
\centering
\includegraphics[width=16.5cm]{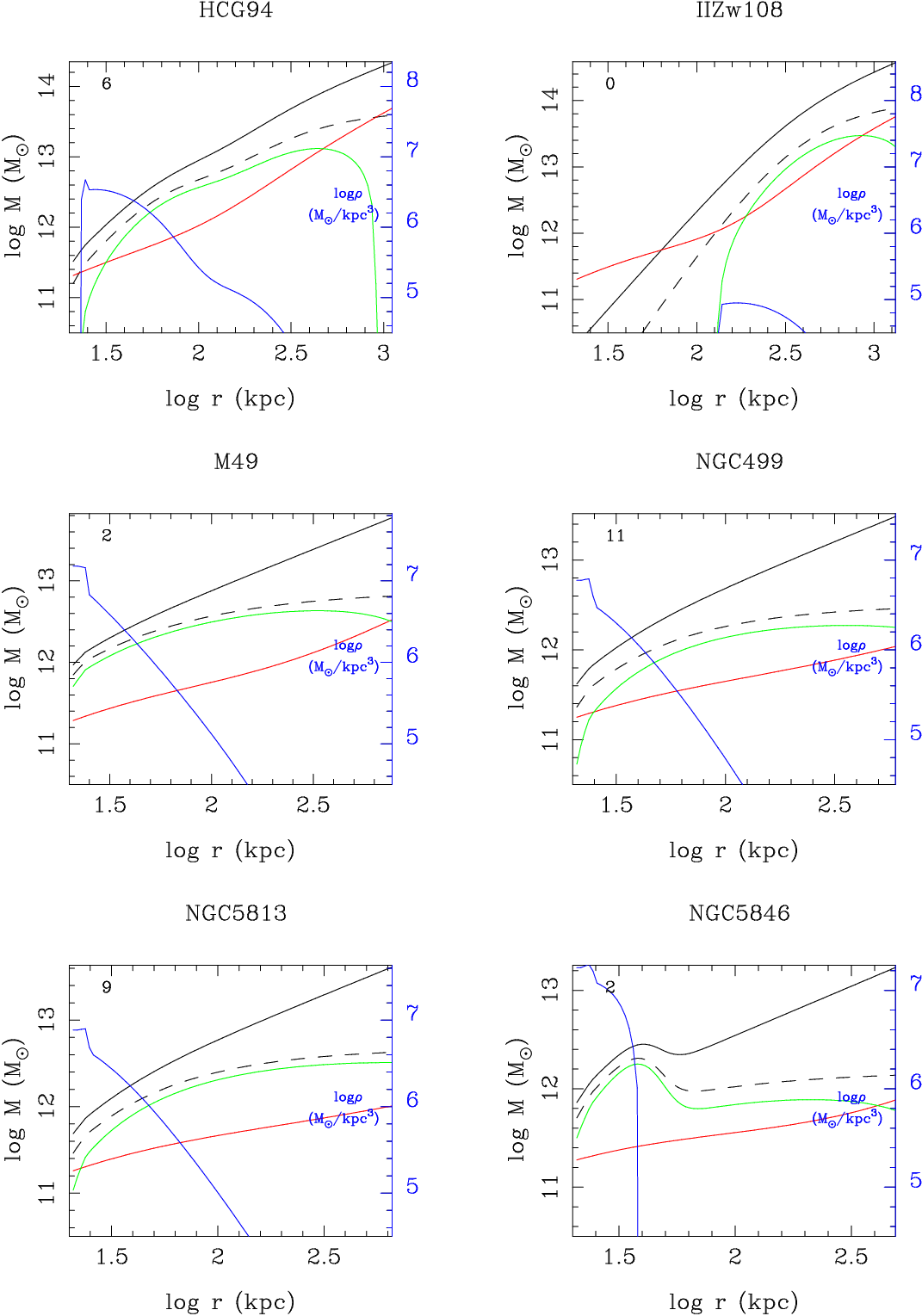}
\caption{ HCG94   , IIZw108 , M49     , NGC499  , NGC5813 , NGc5846}
\end{figure*}

\begin{figure*}
\centering
\includegraphics[width=16.5cm]{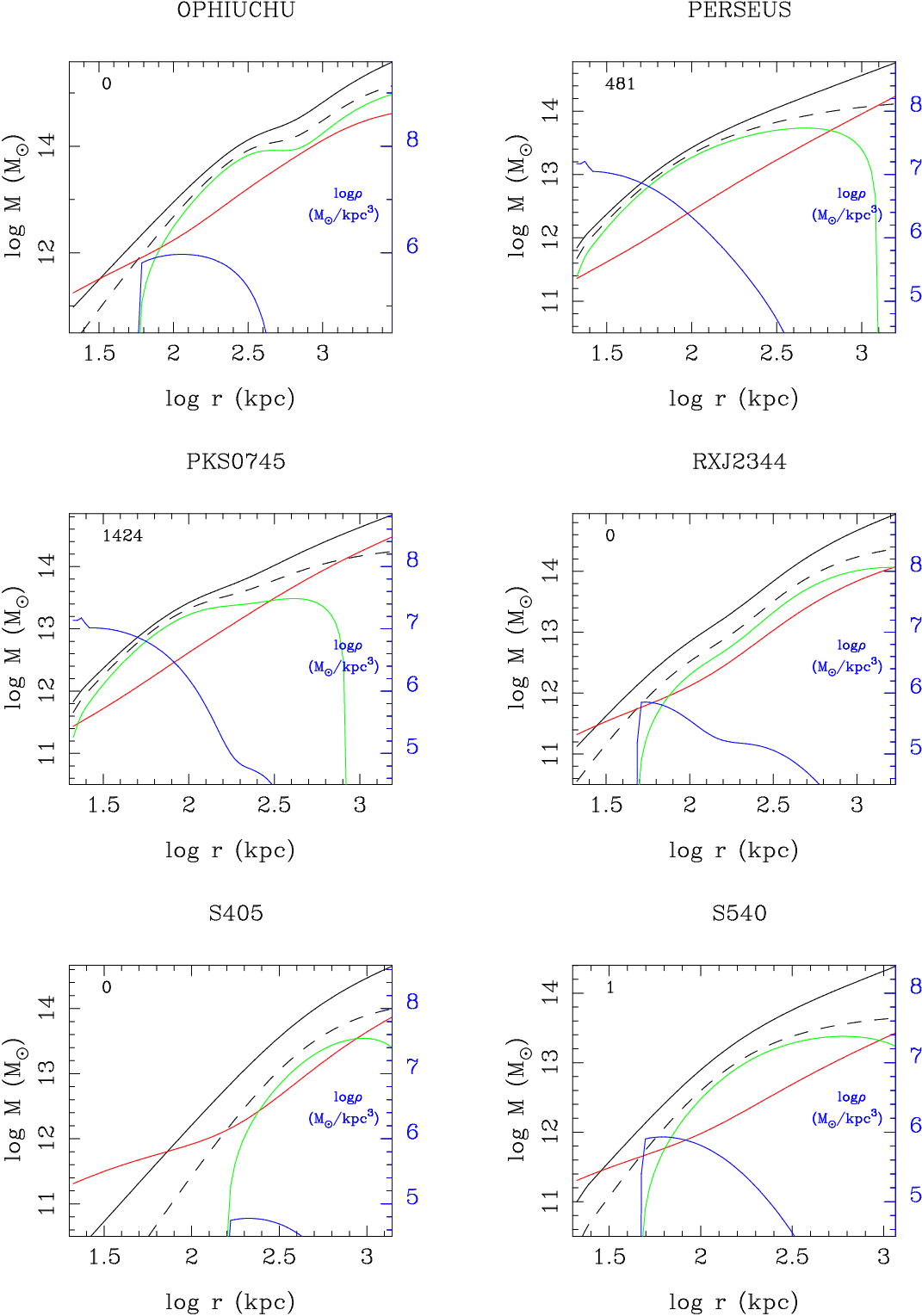}
\caption{ OPHIUCHU, PERSEUS , PKS0745 , RXJ2344 , S405    , S540 }
\end{figure*}

\begin{figure*}
\centering
\includegraphics[width=16.5cm]{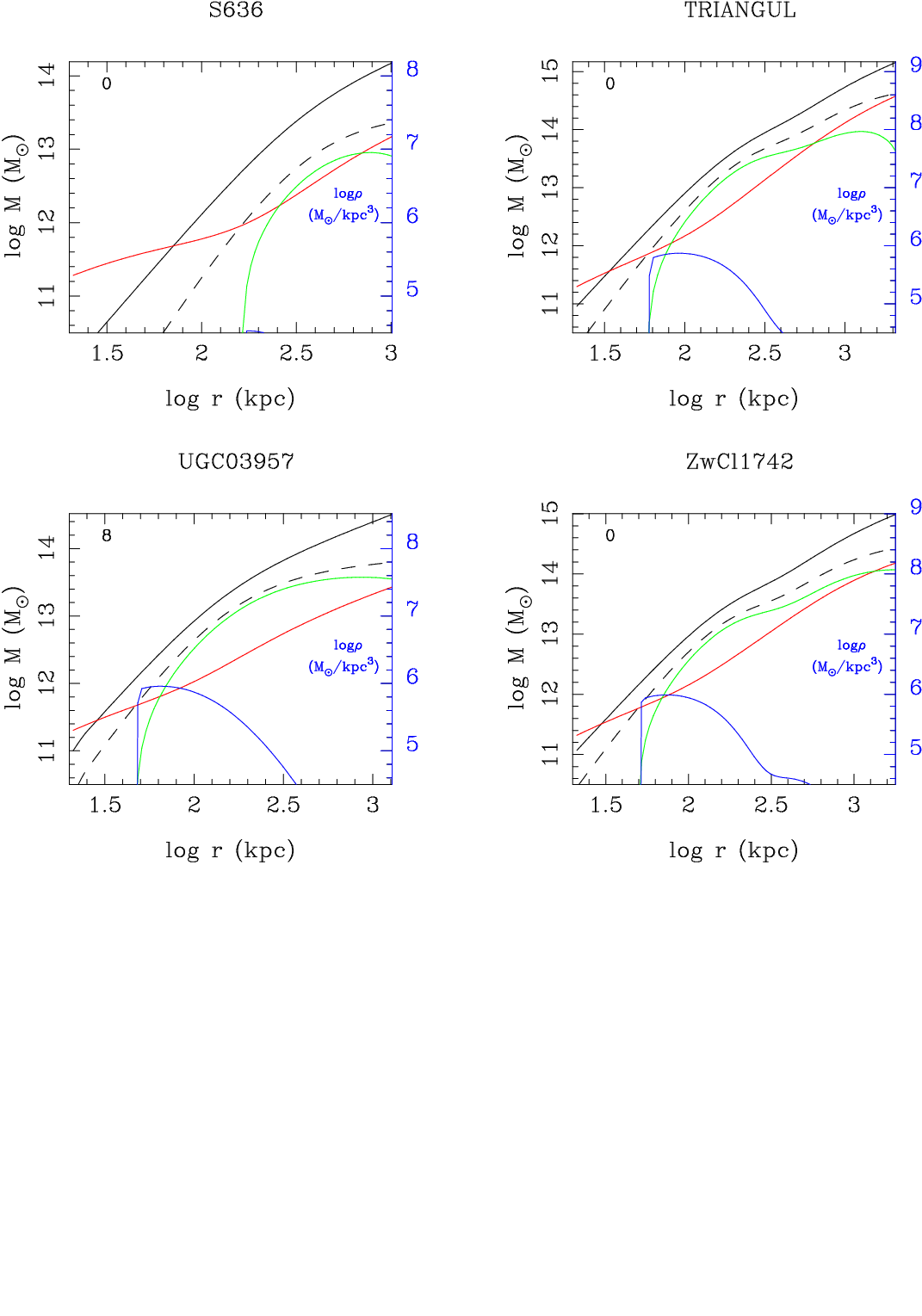}
\caption{S636    , TRIANGUL, UGC03957, ZwCl1742}
\end{figure*}


\end{document}